\documentclass[prd,12pt]{article}
\usepackage{diagbox}

\usepackage{tikz}
\usepackage{amsmath,amssymb,graphicx,multirow,xspace,slashed,array,booktabs}
\usepackage[colorlinks=true,urlcolor=blue,anchorcolor=blue,citecolor=blue,filecolor=blue,linkcolor=blue,menucolor=blue,pagecolor=blue]{hyperref}
\usepackage[compress,numbers]{natbib}
\usepackage{placeins}
\usepackage{subcaption}
\usepackage{booktabs}
\usepackage{blindtext, rotating}
\usepackage{afterpage}
\usepackage{enumitem}
\usepackage{marvosym}
\usepackage{authblk} 
\usepackage{verbatim}
\usepackage{soul} 
\usepackage[normalem]{ulem}
\usepackage{pifont}
\usepackage{booktabs}
\usepackage{bm}
\usepackage{cleveref}
\usepackage{siunitx}
\usepackage{tikz}
\usepackage{tikz-feynman}
\tikzfeynmanset{compat=1.1.0}
\usepackage{cases}
\usepackage{ytableau}
\usepackage{bbm}

\usepackage{floatrow}
\newfloatcommand{capbtabbox}{table}[][\FBwidth]

\usepackage[font=footnotesize,labelfont=bf]{caption}

\usepackage{lineno}

\allowdisplaybreaks

\long\def\symbolfootnote[#1]#2{\begingroup%
\def\thefootnote{\fnsymbol{footnote}}\footnote[#1]{#2}\endgroup}

\allowdisplaybreaks

\makeatletter
	
	\@addtoreset{equation}{section}
\makeatother

\newcommand{\newc}{\newcommand}
\newc{\gsim}{\lower.7ex\hbox{$\;\stackrel{\textstyle>}{\sim}\;$}}
\newc{\lsim}{\lower.7ex\hbox{$\;\stackrel{\textstyle<}{\sim}\;$}}
\newc{\gev}{\,{\rm GeV}}
\newc{\mev}{\,{\rm MeV}}
\newc{\ev}{\,{\rm eV}}
\newc{\kev}{\,{\rm keV}}
\newc{\tev}{\,{\rm TeV}}

\def\tr{\mathop{\rm tr}}

\newc{\mz}{M_Z}
\newc{\mpl}{M_*}
\newc{\mw}{m_{\rm weak}}
\newc{\nr}[1]{N^c_R{}_{#1}}

\usepackage{accents}
\newlength{\dhatheight}

\def\beq{\begin{equation}}
\def\eeq{\end{equation}}
\newcommand{\bea}{\begin{eqnarray}\begin{aligned}}
\newcommand{\eea}{\end{aligned}\end{eqnarray}}
\def\bitem{\begin{itemize}}
\def\eitem{\end{itemize}}

\newcommand{\lmk}{\left(}  
\newcommand{\rmk}{\right)}

\newcommand{\vev}[1]{ \left\langle {#1} \right\rangle }

\renewcommand{\arraystretch}{1.3}

\begin{document}

\baselineskip 0.6cm

\begin{titlepage}

\vspace*{-0.5cm}

\thispagestyle{empty}

\begin{center}

\vskip 1cm

{\LARGE
Specially Embedding a Composite Axion Model
}

\vskip 1.0cm
{Shihwen Hor$^{1,2}$, Yuichiro Nakai$^{1,2}$, Motoo Suzuki$^{3,4,5}$, and Junxuan Xu$^{1,2}$}
\\*[10pt]
$^1${\it \normalsize Tsung-Dao Lee Institute, Shanghai Jiao Tong University, \\
No.~1 Lisuo Road, Pudong New Area, Shanghai, 201210, China} \\*[3pt]
$^2${\it \normalsize School of Physics and Astronomy, Shanghai Jiao Tong University, \\
800 Dongchuan Road, Shanghai, 200240, China} \\*[3pt]
$^3${\it \normalsize SISSA International School for Advanced Studies, \\
Via Bonomea 265, 34136, Trieste, Italy} \\
$^4${\it \normalsize INFN, Sezione di Trieste, Via Valerio 2, 34127, Italy}\\
$^5${\it \normalsize IFPU, Via Beirut 2, 34014 Trieste, Italy}

\vskip 1.0cm

\end{center}

\begin{abstract}

We present a novel framework of the post-inflationary composite axion to address the strong CP problem without the cosmological domain wall problem. Conventional composite axion models lead to the domain wall number greater than one, producing stable axion domain walls that overclose the Universe. We show that by considering a special embedding of the confining gauge group responsible for the composite axion as well as QCD into a larger product gauge group, the domain wall number is essentially set to unity in the ultraviolet (UV) theory. In this setup, small instanton effects associated with the UV gauge dynamics induce a controlled explicit breaking of the residual discrete symmetry, providing a bias term in the axion potential. As a result, the domain walls become unstable and decay sufficiently quickly, while the axion solution to the strong CP problem remains intact. We construct an explicit realization of this framework, identify a viable parameter region and analyze the axion dark matter abundance. Decays of exotic hadrons from the composite dynamics are also investigated. Our special-embedding UV completion renders the domain wall problem in composite axion models cosmologically harmless.

\end{abstract}

\flushbottom

\end{titlepage}

\clearpage
\noindent\makebox[\linewidth]{\rule{\textwidth}{1pt}} 
\setcounter{tocdepth}{2}
\hypersetup{linkcolor=black}
\tableofcontents
\hypersetup{linkcolor=blue}
\noindent\makebox[\linewidth]{\rule{\textwidth}{1pt}} 	

\section{Introduction
\label{sec:introduction}}

The Peccei-Quinn (PQ) mechanism remains the most compelling solution to the strong CP problem
\cite{Peccei:1977hh}. By promoting the QCD vacuum angle to a dynamical field, called the axion, the potential minimum conserves CP symmetry
\cite{Weinberg:1977ma,Wilczek:1977pj}.
Interestingly, this axion field can become a candidate of dark matter (DM)
\cite{Preskill:1982cy,Abbott:1982af,Dine:1982ah}.
Among the various realizations of this idea, composite axion models provide an especially attractive framework in which the axion emerges as a pion-like bound state of a new confining gauge dynamics
\cite{Kim:1984pt,Choi:1985cb} (see Refs.~\cite{Randall:1992ut,Redi:2016esr,Lillard:2017cwx,Lee:2018yak,Gavela:2018paw,Nakai:2021nyf,Vecchi:2021shj,Lee:2021slp,Contino:2021ayn,Nakagawa:2023shi,Cox:2023dou,Nakagawa:2024kcb,Sato:2025rok,Gherghetta:2025fip,Gherghetta:2025kff,Agrawal:2025mke,Azatov:2025mep} for more recent works). A simple realization of such a composite axion is based on a $SU(N)$ gauge theory with vector-like fermions charged under the gauge group. When the confining gauge dynamics generates fermion condensates, the flavor symmetry is spontaneously broken, giving rise to Nambu-Goldstone (NG) bosons, one of which corresponds to the axion field. The unbroken subgroup of the flavor symmetry is gauged and identified as the QCD gauge group, so that the proper axion-gluon coupling is induced through the color anomaly. Since the PQ symmetry is spontaneously broken by the new strong dynamics, composite axion models offer a natural explanation for the hierarchically small PQ breaking scale compared to the Planck scale.

Cosmological consequences of the axion are totally different, depending on whether the PQ symmetry is broken before or after inflation. In the pre-inflationary scenario, the PQ symmetry is broken during or before inflation and never restored afterward. In this case, the axion field is homogenized over our observable Universe, while quantum fluctuations during inflation generate isocurvature perturbations~\cite{Linde:1984ti,Seckel:1985tj,Lyth:1989pb,Turner:1990uz,Linde:1991km}, which are tightly constrained by Cosmic Microwave Background (CMB) observations~\cite{Planck:2018jri}, leading to stringent bounds on the inflation scale. On the other hand, in the post-inflationary scenario, the PQ symmetry is broken after inflation, so that the axion isocurvature problem does not arise. However, this scenario generically leads to the formation of topological defects, such as cosmic strings at the PQ phase transition and domain walls at the QCD phase transition. In composite axion models, the so-called domain wall number $N_{\rm DW}$ associated with the color anomaly is typically larger than one, {\it e.g.} due to fermions charged under both the confining $SU(N)$ and the color $SU(3)_c$ gauge groups. As a result, multiple degenerate vacua appear after QCD confinement, and the associated domain walls form a stable network that eventually dominates the energy density of the Universe. This axion domain wall problem renders the naive post-inflationary realization of composite axion models inconsistent. A straightforward way to eliminate the domain walls is to introduce a bias term that lifts the degeneracy among the vacua and makes the domain walls unstable~\cite{Sikivie:1982qv}. However, an ad hoc introduction of such an explicit PQ breaking term is unsatisfactory and calls for an explanation of its origin.

In this paper, we pursue a new approach to the domain wall problem in composite axion models
(see Refs.~\cite{Hao:2025kcz,Nakagawa:2025suc,Azatov:2025mep} for other solutions). The central idea is to embed the infrared (IR) confining gauge group leading to the composite axion and the ordinary color $SU(3)_c$ into a larger ultraviolet (UV) gauge group in a non-trivial way of {\it special embedding},
which has been recently proposed by the present authors in the context of grand unification equipped with the (elementary) axion
\cite{Hor:2025gxo}. In a special embedding, a group is embedded in a way that does not correspond to the standard Dynkin diagram truncation of a larger group, and therefore the representation content and group-theoretic indices are reorganized in a non-trivial manner. Then, when we identify the IR gauge group in a composite axion model leading to $N_{\rm DW} > 1$ as a special subgroup of a larger UV gauge symmetry, the UV theory can essentially have cosmologically harmless $N_{\rm DW} = 1$. The apparent vacuum degeneracy in the IR theory is lifted by small instanton effects of the UV gauge theory that are not visible in the naive IR description. The small instanton effects can provide a proper bias term in the axion potential to destabilize the domain walls, while the axion retains its role in solving the strong CP problem. We construct an explicit realization of this idea, identify a viable region of the model parameter space and analyze the axion DM abundance.

The rest of the present paper is organized as follows. In section~\ref{sec:composite}, we first describe the original composite axion model and discuss its cosmological problems. Section~\ref{sec:model} then introduces our UV model specially embedding the original model. We estimate small instanton effects on the axion potential in the UV theory. In section~\ref{sec:cosmology}, we discuss the fate of domain walls, find a viable region of the model parameter space and explore the axion DM abundance. Section~\ref{sec:Discussion} is devoted to conclusions and discussion. Appendices summarize the structure of our special embedding and alternative realizations.

\section{Challenges of the composite axion
\label{sec:composite}}

In this section, we begin with a review of the original composite axion model
\cite{Kim:1984pt}
and highlight its cosmological challenges, i.e. the axion domain wall problem and the presence of exotic hadrons as stable charged massive particles (CHAMPs). The domain wall issue will be addressed through a special embedding as discussed in later sections. We here explore a solution to the CHAMP problem by introducing exotic hadron interactions with the SM particles, allowing their decays.

\subsection{Original model~\label{sec:rev}}

\begin{table}
    \centering
    \renewcommand{\arraystretch}{1.3}
    \setlength{\tabcolsep}{1.3pt}
    \begin{tabular}{|c|c|c|c|c|c|}
        \hline
       \, Fields \, & \, $[SU(3)_c]$ \, & \, $[SU(N)]$ \, & \, $U(1)_{\rm PQ}$ \, & \, $[U(1)_Y]$ \,  \\ \hline
       $Q$&    $\mathbf{3}$ & $\mathbf{N}$ &  1 & $a$\\ 
       $\eta$ & $\mathbf{1}$ & $\mathbf{N}$ &  $-3$ & $b$ \\ \hline
         $\bar Q$ &  $\bar{\mathbf{3}}$ & $\mathbf{\overline N}$  & $0$ & $-a$\\ 
        $\bar\eta$ &   $\mathbf{1}$ & $\mathbf{\overline N}$ & $0$  &  $-b$\\ \hline
    \end{tabular}
    \vspace{3mm}
    \caption{Matter content for the original composite axion model.
    }
    \label{tab:org}
\end{table}

The original composite axion model
is based on a $SU(N)$ gauge theory with four pairs of vector-like fermions in the fundamental and anti-fundamental representations, respecting a global $SU(4)_L \times SU(4)_R$ flavor symmetry
whose diagonal $SU(4)_V$ subgroup is partly gauged and identified with the ordinary color $SU(3)_c$ gauge group. The global $U(1)_{\rm PQ}$ symmetry is embedded in the $SU(4)_L \times SU(4)_R$ flavor symmetry and carries anomaly coefficients,
\[
\mathcal{A}(U(1)_{\rm PQ}-SU(3)_c^2 ) = N \, , \qquad
\mathcal{A}(U(1)_{\rm PQ}-SU(N)^2) = 0 \, ,
\]
ensuring the proper axion coupling to QCD while keeping anomaly-free under $SU(N)$. The matter content and charge assignments are summarized in Table~\ref{tab:org}, where $[G]$ represents a gauge symmetry while $G$ a global symmetry.\footnote{In the original model, the extra quarks are neutral under the $U(1)_Y$ hypercharge gauge group.}

The $SU(N)$ gauge theory is asymptotically free and confines at low-energies. 
Below the $SU(N)$ confinement scale, fermion condensates break the flavor symmetry as
\[
SU(4)_L \times SU(4)_R \to SU(4)_V \ ,
\]
generating fifteen pseudo-Nambu-Goldstone Bosons (pNGBs) which are in the fifteen representation of $SU(4)$, and transform under $SU(3)_c$ as
\begin{align}
{\bf 15}\to 
 \mathbf{1} \oplus \mathbf{3} \oplus \mathbf{\bar 3} \oplus \mathbf{8}\ ,
\end{align}
where the singlet corresponds to the axion.
The colored states obtain masses in a similar manner to the charged pions, estimated as~\cite{Choi:1985cb},
\begin{align}
    m_{\bf r}^2\approx \frac{g_c^2(F_a)}{16\pi^2}C_2({\bf r})\Lambda_{\rm conf}^2 \ ,
\end{align}
where $g_c$ is the $SU(3)_c$ gauge coupling constant, $F_a$ denotes the decay constant of the pNGBs, $C_2({\bf r})$ is the quadratic Casimir for the representation ${\bf r}$ and $\Lambda_{\rm conf}$ is the confinement scale of $SU(N)$.
While there exits another pNGB associated with the anomalous $U(1)_A$ symmetry where $Q$ and $\eta$ have the same charges, it acquires a mass from the $SU(N)$ strong dynamics, analogous to the $\eta'$ meson in QCD.

The anomaly coefficient $\mathcal{A}(U(1)_{\rm PQ}-SU(3)_c^2) = N$ gives rise to an anomaly-free subgroup $\mathbb{Z}_N \subset U(1)_{\rm PQ}$. 
As a result, the axion potential possesses $N$ degenerate vacua, leading to the formation of stable domain walls around the QCD phase transition. 
We address this issue by lifting the vacuum degeneracy through small-instanton effects of a UV gauge theory, as discussed in later sections.

Another cosmological issue is the presence of stable $SU(N)$ hadrons charged under $SU(3)_c$. 
These hadrons couple with SM quarks around the QCD phase transition and acquire electromagnetic charges. 
Such charged massive particles (CHAMPs) are strongly constrained by various experiments. 
In the following, we analyze the hadron spectrum and then discuss their possible decays into the SM sector.

\subsubsection{Baryons at $N=3$ and mesons}
For $N=3$, the baryons are constructed as
\begin{align}
  B \sim \Psi_a \Psi_b \Psi_c \, \epsilon^{abc} \ , \qquad \bar B \sim \bar\Psi^a \bar\Psi^b \bar\Psi^c \, \epsilon_{abc} \ ,
\end{align}
where $\Psi \sim Q,\eta$ and $\bar\Psi \sim \bar Q,\bar\eta$,
and $\epsilon^{abc}$ with the dark $SU(3)$ gauge indices $a,b,c = 1,2,3$ is a totally antisymmetric tensor.  
In addition, $\Psi$ carries a $SU(4)_L$ flavor index as well as a spin index. 
Since $\Psi$ is a Grassmann-valued fermionic field, the total wavefunction must be antisymmetric under the exchange of fermions. 
Given the antisymmetric contraction of gauge indices, the combined spin and flavor wavefunction must therefore be symmetric in order to obtain a nonvanishing state.
To find the corresponding flavor and spin states, we 
consider the embedding,
\begin{align}
    SU(4)\times SU(2)\subset SU(8)\ ,
\end{align}
where $SU(4)$ and $SU(2)$ are for the flavor symmetry and spin, respectively.%
\footnote{This is the same strategy to find the octet and decuplet baryons with $SU(6)$ in the quark model~\cite{Georgi:1982jb}.}
Each quark is embedded into the fundamental representation of the $SU(8)$, which is decomposed into the bifundamental  $(\mathbf{4},\mathbf{2})$  representation of the $SU(4)\times SU(2)$.
In the $SU(8)$, the totally symmetric representation for the flavor and spin state of the baryon $B$ corresponds to
\begin{align}
 \mathbf{120} =   \ytableaushort{\,\,\,}\ ,
\end{align}
where the right-hand side denotes the corresponding Young tableau for $SU(8)$.
This is decomposed into 
\begin{align}
    \mathbf{120} \to (\mathbf{\overline{20}''},\mathbf{4})\oplus(\mathbf{\overline{20}},\mathbf{2})\ ,
\end{align}
in terms of $SU(4)\times SU(2)$.
On the right-hand side, two components have different representations under $SU(4)$ with Young tableaux,
\begin{align}
     \ytableaushort{\,\,\,}\ , \qquad \qquad 
    \ytableaushort{\,\, ,\,}\ ,
\end{align}
respectively. 
Therefore, we find that there exist a 20-plet spin 3/2 baryon and a 20-plet spin 1/2 baryon.
A part of $SU(4)$ is gauged and identified with the $SU(3)_c$. In terms of $SU(3)_c$, we have
\begin{align}
(\mathbf{\overline{20}''},\mathbf{4}) &\to (\mathbf{10}_3,\mathbf{4})\oplus (\mathbf{\bar 6}_{-1},\mathbf{4})\oplus (\mathbf{3}_{-5},\mathbf{4})\oplus (\mathbf{1}_{-9},\mathbf{4})\ , \\[1ex]
(\mathbf{\overline{20}},\mathbf{2}) &\to (\mathbf{8}_3,\mathbf{2})\oplus (\mathbf{\bar{6}}_{-1},\mathbf{2})\oplus (\mathbf{\bar{3}}_{-1},\mathbf{2})\oplus (\mathbf{3}_{-5},\mathbf{2})\ ,
\end{align}
where the subscript denotes the $U(1)$ of $U(1)\times SU(3) \subset SU(4)$.
The anti-baryons can be also studied in the same way.

On the other hand, for mesons, we have
\begin{align}
    \mathbf{8}\otimes \mathbf{\bar 8}= \mathbf{1} \oplus \mathbf{63}\ ,
\end{align}
and the decomposition of the $SU(8)$ adjoint representation in terms of $SU(4)\times SU(2)$ is given by
\begin{align}
    \mathbf{63}\to (\mathbf{15},\mathbf{3})\oplus (\mathbf{1},\mathbf{3})\oplus (\mathbf{15},\mathbf{1})\ .
\end{align}
In terms of $SU(3)_c$, we find
\begin{align}
   (\mathbf{15},\mathbf{3}) &\to  (\mathbf{3}_4,\mathbf{3})\oplus (\mathbf{8}_0,\mathbf{3})\oplus (\mathbf{1}_0,\mathbf{3})\oplus (\mathbf{\bar 3}_{-4},\mathbf{3}) \ , \\[1ex]
    (\mathbf{15},\mathbf{1}) &\to  (\mathbf{3}_4,\mathbf{1})\oplus (\mathbf{8}_0,\mathbf{1})\oplus (\mathbf{1}_0,\mathbf{1})\oplus (\mathbf{\bar 3}_{-4},\mathbf{1})\ .
\end{align}
The spectrum of the spin zero mesons is summarized in Table~\ref{tab:meson_0}.
The meson spectrum is the same for other $N$ of $SU(N)$.

\begin{table}[t!]
    \centering
    \renewcommand{\arraystretch}{1.5}
    \begin{tabular}{|c|c|c|c|c|c|c|c|c|c|}
        \hline
      Fields &  $[SU(3)_c]$ & $[U(1)_Y]$  
         \\ \hline
    $\Phi_3^{2/3}$ & $\mathbf{3}$  & $a-b=2/3$ 
     \\ 
     \hline
     $\phi_8$ & $\mathbf{8}$ & $0$ 
      \\ 
      \hline
     $a$ &   $\mathbf{1}$ & $0$  
     \\ \hline
     $ a_\eta$ & $\mathbf{1}$ & $0$ 
     \\ 
     \hline
     \end{tabular}
     \vspace{3mm}
    \caption{Spin 0 meson spectrum.
    Here, $\Phi_3^{2/3}$ denotes the complex scalar field, while the others are real scalar fields.
    }
    \label{tab:meson_0}
\end{table}

\subsubsection{Baryons at $N=4$}

We now analyze the baryon spectrum for the case with $N=4$. As in the case of $N=3$, the total baryon wavefunction must
be antisymmetric under the exchange of constituent fermions. Given the antisymmetric contraction of gauge indices, the combined spin and flavor wavefunction must be symmetric. Considering the embedding of $SU(4)\times SU(2)\subset SU(8)$,
for $N=4$, the totally symmetric representation of $SU(8)$ corresponds to
\begin{align}
 \mathbf{330} =   \ytableaushort{\,\,\,\,}\ ,
\end{align}
which is decomposed in terms of $SU(4)\times SU(2)$ as 
\begin{align}
    \mathbf{330} \to  (\mathbf{35},\mathbf{5})\oplus (\mathbf{45},\mathbf{3})\oplus (\mathbf{20}',\mathbf{1})\ .
\end{align}
In terms of $SU(3)_c$, we find
\begin{align}
 (\mathbf{35},\mathbf{5}) &\to (\mathbf{15}'_4,\mathbf{5})\oplus(\mathbf{10}_0,\mathbf{5})\oplus (\mathbf{\bar 6}_{-4},\mathbf{5})\oplus (\mathbf{3}_{-8},\mathbf{5})\oplus (\mathbf{1}_{-12},\mathbf{5})\ ,\\[1ex]
  (\mathbf{45},\mathbf{3}) &\to (\mathbf{15}_4,\mathbf{3})\oplus(\mathbf{10}_0,\mathbf{3})\oplus ( \mathbf{8}_{0},\mathbf{3})\oplus (\mathbf{\bar 6}_{-4},\mathbf{3})\oplus (\mathbf{\bar 3}_{-4},\mathbf{3})\oplus (\mathbf{3}_{-8},\mathbf{3})\ ,\\[1ex]
   (\mathbf{20}',\mathbf{1}) &\to (\mathbf{6}_4,\mathbf{1})\oplus(\mathbf{8}_0,\mathbf{1})\oplus (\mathbf{\bar 6}_{-4},\mathbf{1})\ ,
\end{align}
where the subscript also denotes the $U(1)$ of $U(1)\times SU(3) \subset SU(4)$.
The spin 0 scalar baryon spectrum is summarized in Table~\ref{tab:baryons_N4_0}.

\begin{table}[t!]
    \centering
    \renewcommand{\arraystretch}{1.5}
    \begin{tabular}{|c|c|c|c|c|}
        \hline
         Fields & $[SU(3)_c]$ & $[U(1)_Y]$   \\ \hline
        $B_6 \in QQQQ$ & $\mathbf{6}$  & $4a=2/3$ \\ 
     \hline
     $B_8 \in QQQ\eta$ & $\mathbf{8}$ & $3a+b=0$ \\ \hline
     $B_{\bar 6} \in QQ\eta\eta$ & $\mathbf{\bar 6}$ &  $2a+2b=-2/3$ \\ \hline
     \end{tabular}
     \vspace{3mm}
    \caption{Spin 0 baryons for $N=4$.
    }
    \label{tab:baryons_N4_0}
\end{table}

\subsection{Exotic hadron decays}

As we have seen, the low-energy spectrum of the composite axion model contains a variety of baryons and mesons, including spin-0 pNGBs and scalar baryons. Some of them are stable, as no interactions allow them to decay into lighter states unless we introduce extra higher-dimensional operators suppressed by a cutoff scale. 
Such stable exotic hadrons pose a cosmological problem. Their annihilation cross sections are insufficient to efficiently deplete their number densities due to their large masses, $m \gg 10^5~\mathrm{GeV}$~\cite{Griest:1989wd}. Furthermore, hadrons charged under $SU(3)_c$ bind with SM quarks at the QCD phase transition, forming electromagnetically charged bound states. These charged massive particles (CHAMPs) are strongly constrained by cosmological and astrophysical observations (see e.g. Ref.~\cite{Dunsky:2018mqs} for a recent analysis) and cannot constitute viable dark matter candidates.
To address this exotic relic problem, the following two general solutions exist:
\begin{itemize}
    \item Introduction of decay channels to the SM sector via higher-dimensional operators.  
    \item Dilution of the relic abundance via e.g. a secondary mini-inflation~\cite{Azatov:2025mep}.%
    \footnote{The scenario can be applied to our model.}
\end{itemize}
Here, we focus on the first option, which has received tiny attention, likely due to model-building complexity, but is essential for a complete cosmological treatment. We take the case of $N=4$ as an example and consider decay mechanisms for spin-0 mesons (pNGBs) and scalar baryons.


Let us introduce higher-dimensional interactions suppressed by a cutoff scale $\Lambda_{\rm cutoff}$ that couple exotic mesons to SM fields. For instance, for a pNGB $\Phi_3^{2/3}$ in the fundamental representation of $SU(3)_c$ (see Tab.~\ref{tab:meson_0}), one can write e.g.

\begin{align}
   \mathcal{L} \sim \frac{1}{\Lambda_{\rm cutoff}^2}\,
   \bar Q\eta\, \bar d \bar d \,,
   \label{eq:decay_1}
\end{align}
where $\bar d$ denotes the $SU(2)_L$ singlet down-type quarks, and the hypercharge assignments ensure gauge invariance.
With an appropriate assignment of PQ charges to the SM fields, the $U(1)_{\rm PQ}$ symmetry can be preserved 
while the anomaly coefficient is unaffected. 
Other mesons decay through cascade decays. For example, an octet meson $\phi_8$ can decay via
\begin{align}
    \phi_8 \to \Phi_3^{2/3} +\Phi_{\bar 3}^{-2/3} + a \to \bar d + \bar d + \Phi_{\bar 3}^{-2/3} + a \,,
\end{align}
where the intermediate $\Phi_3^{2/3}$ is off-shell because the mass ratio,
\begin{align}
    \frac{m_{\phi_8}}{m_{\Phi_3}} \sim \sqrt{\frac{C_2({\bf 8})}{C_2({\bf 3})}} \simeq \frac{3}{2}  \ , 
\end{align}
prevents an on-shell decay. The heavier meson $a_\eta$ decays via couplings to topological terms of $SU(N)$.

For scalar baryons, their decays require (dark) baryon-number-violating operators. An example operator for the octet baryon $B_8^i$ ($i=1,...,8$) is
\begin{align}
  \mathcal{L}\sim \frac{(QQQ\eta)^i}{\Lambda_{\rm cutoff}^6} 
   H q T^i \bar u \,,
   \label{eq:decay_2}
\end{align}
where $H$ is the SM Higgs doublet, $q$ the quark doublet, $T^i$ the $SU(3)_c$ generator, and $\bar u$ the up-type quark singlet.  
The baryon in the symmetric representation $B_6$ can decay through e.g.
\begin{align}
    B_6 \to B_8 + \Phi_3^{2/3} \,,
\end{align}
where $B_8$ is off-shell. $\bar B_6$ decays analogously due to mass mixing. Anti-baryons decay in a similar manner.  

Requiring sufficiently rapid decays, the cutoff scales must be low enough to ensure that exotic meson and baryon decay before they dominate the energy density of the Universe.
We do not attempt to specify UV completion responsible for the higher-dimensional operators.%
\footnote{
A UV origin for the operators in Eqs.~\eqref{eq:decay_1} and \eqref{eq:decay_2} can be obtained by introducing heavy scalar mediators at the cutoff scale. For example, a scalar $\Phi$ with mass $M_\Phi \sim \Lambda_{\rm cutoff}$ and couplings $y_1 \bar Q \eta \Phi + y_2 \Phi^\ast \bar d \bar d$ generates the dimension-6 operator in Eq.~\eqref{eq:decay_1} upon integrating out $\Phi$ at tree level. A similar construction with additional heavy scalars and appropriate gauge charges reproduces Eq.~\eqref{eq:decay_2}. All mediators are unstable due to the same renormalizable couplings, so no stable exotic relics are introduced. A more complete embedding (e.g. composite or warped/conformal scenarios) may provide a dynamical origin for the scalar mass scale and a possible explanation of the hierarchy, and is left for future work.
} 
Generally, the introduction of intermediately heavy scalar fields for such UV completion generates a hierarchy problem, threatening the motivation of composite axion models.
The heavy states at the cutoff scales may affect the running of gauge couplings and modify small-instanton contributions,
which we will not get into further in the present study.

\section{Specially embedded composite axion  
\label{sec:model}}

As discussed in the previous section, the domain wall problem remains a significant challenge for the composite axion model. In this section, we propose a specially embedded composite axion model to address this issue. A new approach to the domain wall problem via special embedding has been introduced in Ref.~\cite{Hor:2025gxo}.
In this construction, the IR gauge group is non-trivially embedded into a larger UV gauge group.
This setup can make it possible to obtain a Dynkin index of a PQ-charged fermion representation for the IR gauge group different from that for the UV gauge group, and consequently the domain wall number differs among the UV and IR theories. Details of special embedding are summarized in Appendix~\ref{sec:special}. 

\subsection{The setup}

We consider a $SU(4N)\times SU(3)_2\times SU(N)_2\times U(1)_X$ gauge symmetry for our UV theory.\footnote{We omit the $SU(2)_L$ gauge symmetry of the SM since it is irrelevant to the following discussion. The full gauge symmetry of the model should be $SU(4N)\times SU(3)_2\times SU(N)_2\times U(1)_X\times SU(2)_L$.}
The matter content of the model besides from the SM particles and the corresponding charges under the symmetry groups are summarized in Table~\ref{tab:special2}. 
We introduce Weyl fermions $\Psi,~\bar\Psi$, transforming as fundamental and anti-fundamental representations under $SU(4N)$, and complex scalar fields $\Phi$, $\Phi'$, and $\Phi''$, whose vacuum expectation values (VEVs) spontaneously break the UV gauge symmetry into that of the original composite axion model. The SM quarks are charged under the $SU(3)_2 \times U(1)_X$ gauge symmetry, which will make them charged under the SM $SU(3)_c \times U(1)_Y$ after the gauge symmetry breaking. The global symmetry of our UV model is given by 
$SU(4N)_L \times SU(4N)_R \times U(1)_A \times U(1)_B$.
We can consider that the diagonal subgroup of the $SU(4N)_L\times SU(4N)_R$ flavor symmetry group is gauged and identified as $[SU(4N)]$.

\begin{table}
    \centering
    \renewcommand{\arraystretch}{1.5}
    \setlength{\tabcolsep}{1.5pt}
    \begin{tabular}{|c|c|c|c|c|c|c|c|c|c|}
        \hline
       Fields &  $[SU(4N)]$  & $[SU(3)_2]$ & $[SU(N)_2]$ & $[U(1)_{X}]$ & $U(1)_{A}$  & $U(1)_{B}$  & $SU(4N)_{L}$ & $SU(4N)_{R}$  \\ \hline
       $\Psi$&    $\mathbf{4N}$ & $\mathbf{1}$ & $\mathbf{1}$ & $x$ & 1  & 1 &  $\mathbf{4N}$ & $\mathbf{1}$ \\ \hline
       $\bar\Psi$ & $\overline{\mathbf{4N}}$ & $\mathbf{1}$ & $\mathbf{1}$ & $-x$ & 0 & $-1$  & $\mathbf{1}$ & $\overline{\mathbf{4N}}$ \\ \hline
       $\Phi$ & $\mathbf{(4N)^2-1}$ & $\mathbf{8}$  &  $\mathbf{1}$ & 0 & 0 & 0 &  $\mathbf{1}$ & $\mathbf{1}$ \\ \hline
       $\Phi'$ & $\mathbf{(4N)^2-1}$ & $\mathbf{1}$ & $\mathbf{N^2-1}$ & 0 & 0 & 0 &  $\mathbf{1}$ & $\mathbf{1}$ \\ \hline
       $\Phi''$ & $\mathbf{4N}$ & $\mathbf{1}$ & $\overline{\mathbf{N}}$ & $3z$ & 0 & 0 &  $\mathbf{1}$ & $\mathbf{1}$ \\ \hline
    \end{tabular}
    \vspace{3mm}
    \caption{Matter content for the specially embedded composite axion model.
    }
    \label{tab:special2}
\end{table}

The $SU(4N)$ gauge group contains a special subgroup of $SU(4)_1\times SU(N)_1$. Appendix~\ref{sec:special} summarizes the special embedding, $SU(4)_1\times SU(N)_1\subset SU(4N)$.
The scalar fields $\Phi$, $\Phi'$, and $\Phi''$
can be decomposed under $SU(4N) \times SU(3)_2 \times SU(N)_2 \times U(1)_X \to [ SU(4)_1 \times SU(N)_1 ]_{4N} \times SU(3)_2 \times SU(N)_2 \times U(1)_X$ as
\begin{align}
    \Phi &=(\mathbf{(4N)^2-1} ,\, \mathbf{8} ,\, \mathbf{1}, \, 0) \nonumber\\
    &\to (\mathbf{15} ,\, \mathbf{N^2-1} ,\, \mathbf{8} ,\, \mathbf{1}, \, 0) \oplus (\mathbf{15} ,\, \mathbf{1} ,\, \mathbf{8} ,\, \mathbf{1}, \, 0) \oplus (\mathbf{1} ,\, \mathbf{N^2-1} ,\, \mathbf{8} ,\, \mathbf{1}, \, 0)~,\nonumber\\[1ex]
    \Phi' &=(\mathbf{(4N)^2-1}, \, \mathbf{1} ,\, \mathbf{N^2-1},\, 0) \nonumber\\
    &\to (\mathbf{15} ,\, \mathbf{N^2-1} ,\, \mathbf{1} ,\, \mathbf{N^2-1}, \, 0) \oplus (\mathbf{15} ,\, \mathbf{1} ,\, \mathbf{1} ,\, \mathbf{N^2-1}, \, 0) \oplus (\mathbf{1} ,\, \mathbf{N^2-1} ,\, \mathbf{1} ,\, \mathbf{N^2-1}, \, 0) ~,\nonumber\\[1ex]
    \Phi'' &=(\mathbf{4N}, \, \mathbf{1} ,\, \overline{\mathbf{N}},\, 3z) 
    \to (\mathbf{4}, \, \mathbf{N}, \, \mathbf{1} ,\, \overline{\mathbf{N}},\, 3z)  ~. 
\end{align}
Note that $\Phi'$ contains the bi-adjoint representation of $SU(N)_1\times SU(N)_2$, while $\Phi$ contains the bi-adjoint representation of $[SU(3)_1\subset SU(4)_1]\times SU(3)_2$.
The bi-adjoint components obtain VEVs and spontaneously break the gauge groups into the corresponding diagonal subgroups, respectively.
For example, $\Phi\supset (\mathbf{15} ,\, \mathbf{1} ,\, \mathbf{8} ,\, \mathbf{1}, \, 0)$ develops a VEV at the diagonal direction of the adjoint representations $\mathbf{15}$ and $\mathbf{8}$. 
This component of $\Phi$ can be described as
\begin{align}
    (\Phi^A_B)^c_d=(\Phi)^I_i (T_{SU(4)}^I)^A_B(T_{SU(3)}^i)^c_d~,
    \label{eq:phi}
\end{align}
where $(T_{SU(4)}^I)^A_B$ ($I=1,...,15$) are the $SU(4)$ generators with the (anti-)fundamental representation indices $A,B=1,...,4$, and $(T_{SU(3)}^i)^c_d$ ($i=1,...,8$) are the $SU(3)$ generators with the (anti-)fundamental indices $c,d=1,2,3$.
Note that the generators of $SU(3)_1\subset SU(4)_1$ can be expressed as 
\begin{equation}
    T_{SU(4)}^{I=1,\cdots,8} = \begin{pmatrix} T_{SU(3)}^{i=I} & 0\\ 0& 0 \end{pmatrix}~.
\end{equation}
The right-hand side of Eq.~\eqref{eq:phi} indicates that this component can be written as a $15\times 8$ matrix $(\Phi)^I_i$, and it develops a VEV at the diagonal direction as
\begin{align}
    \langle\Phi^I_i\rangle = \begin{pmatrix}
        v{\bf 1}_{8\times 8} \\ {\bf 0}_{7\times 8}
    \end{pmatrix}_{15\times 8}~,
\end{align}
which breaks the symmetry $SU(4)_1 (\supset SU(3)_1 \times U(1)_Z) \times SU(3)_2\to SU(3)_V \times U(1)_Z$. 
The remaining symmetry $U(1)_Z\subset SU(4)_1$ besides form $SU(3)_V$ can be found by checking the commutation relation. The $SU(4)_1$ generator $(T^{15}_{SU(4)})^A_B=\frac{1}{2\sqrt{6}}{\rm diag}(1,1,1,-3)$ commutes with all the $SU(3)_1\subset SU(4)_1$ generators, and hence, the VEV $\langle\Phi\rangle$ does not break the corresponding $U(1)_Z\subset SU(4)_1$ symmetry.

Similarly, we can achieve the symmetry breaking of $SU(N)_1\times SU(N)_2\to SU(N)_V$ by a VEV at the diagonal direction of $(\mathbf{1} ,\, \mathbf{N^2-1} ,\, \mathbf{1} ,\, \mathbf{N^2-1},\, 0)\subset \Phi'$, which is described as $(\Phi')^J_K$, where $J,~K=1,...,(N^2-1)$ is the $SU(N)$ adjoint index,
\begin{align}
    \langle (\Phi')^J_K\rangle
  =v'\delta^J_K\ .
\end{align}
In summary, the symmetry-breaking pattern can be described as
\begin{align}
    & SU(4N) \times SU(3)_2 \times SU(N)_2 \times U(1)_X \nonumber \\
    &\qquad \qquad \qquad \qquad \xrightarrow{\langle\Phi\rangle} SU(3)_V \times U(1)_Z\times SU(N)_{4N} \times SU(N)_2 \times U(1)_X ~, \\[2ex]
    & SU(4N) \times SU(3)_2 \times SU(N)_2 \times U(1)_X \nonumber \\
    &\qquad \qquad \qquad \qquad \xrightarrow{\langle\Phi'\rangle} SU(N)_V \times SU(4)_{4N} \times SU(3)_2\times U(1)_X ~.
\end{align}
Considering both VEVs at the same time, we achieve the symmetry breaking,
\begin{equation}
    SU(4N) \times SU(3)_2 \times SU(N)_2 \times U(1)_X \to SU(N)_V \times SU(3)_V\times U(1)_Z \times U(1)_X~,
\end{equation}
where $SU(3)_V$ is identified as $SU(3)_c$.

The SM hypercharge gauge group can be obtained from a linear combination of the $U(1)_Z$ and $U(1)_X$.
Since the Higgs field $\Phi''$ is charged under the $U(1)_X$, a VEV of the component $(\mathbf{1},\mathbf{1},-3,3z) \subset \Phi''$ in terms of $SU(N)_V \times SU(3)_c \times U(1)_Z \times U(1)_X$,
\begin{align}
    & \langle\Phi''\rangle =\langle (\mathbf{1},\mathbf{1},-3,3z) \rangle =v'' ~,
\end{align}
breaks $U(1)_Z  \times U(1)_X $ into such a linear combination,
\begin{align}
    & SU(N)_V\times SU(3)_c \times U(1)_Z \times U(1)_X \xrightarrow{\langle\Phi''\rangle} SU(N)_V\times SU(3)_c \times U(1)_Y ~.
\end{align}
The SM hypercharge is defined as $Q_Y=\sqrt{24}\, z\,T^{15}+Q_X=z\, Q_Z+Q_X$, where $T^{15}=\frac{1}{2\sqrt{6}}{\rm diag}(1,1,1,-3)$ denotes the $SU(4)_1$ generator.
Combining all the VEVs, we finally obtain the gauge symmetry of the original composite axion model,
\begin{equation}
        SU(4N) \times SU(3)_2 \times SU(N)_2 \times U(1)_X \to SU(N)_V \times SU(3)_c\times U(1)_Y~.
\end{equation}
Note that we have focused on the symmetry breaking pattern to realize the composite axion model
by assuming the form of the Higgs VEVs.
The exact form of the Lagrangian and the shape of the multiple Higgs potential that can trigger the symmetry breaking of our model lie beyond the scope of the present work.
The scalar field setup in Table~\ref{tab:special2} is one possibility to achieve the special embedding using bi-adjoint representations. There are also some alternative setups to obtain the similar special embedding structure, which are described in Appendix~\ref{sec:alternative}.

Under $SU(N)_V\times SU(3)_c \times U(1)_Y$, the fermions $\Psi$ and $\bar\Psi$ transform as
\begin{align}
    \Psi:(\mathbf{N},\mathbf{3},z+x)\oplus (\mathbf{N},\mathbf{1},-3z+x)~,\quad \bar \Psi:(\overline{\mathbf{N}},\overline{\mathbf{3}},-z-x)\oplus (\overline{\mathbf{N}},\mathbf{1},3z-x)\ ,
\end{align}
which is the same as the original composite axion model in Section~\ref{sec:rev} after identifying the two components of $\Psi ~(\bar\Psi)$ with $Q,\eta ~(\bar Q,\bar \eta)$ respectively and the $U(1)_Y$ charges $a=z+x,~b=-3z+x$. 
For example, for $N=4$ case, we obtain $a=1/6,~b=-1/2$ and $z=1/6,~ x=0$.
As explained in Section~\ref{sec:rev}, the $SU(4)_L \times SU(4)_R$ flavor symmetry, under which $\Psi,\bar\Psi$ are charged, contains the global $U(1)_{\rm PQ}$ symmetry.
The axion appears in the phase of the quark condensates after the $SU(N)_V$ confinement,
\begin{align}
    \vev{\Psi\bar{\Psi}}\simeq \Lambda_{\rm conf}^3 \Sigma(x) = \Lambda_{\rm conf}^3 \exp\left( \frac{2i}{F_a}\begin{pmatrix}
		\pi_8(x) + \frac{a(x)}{2\sqrt{6}}\mathbf{1}_{3\times 3} & K(x) \\[4pt]
		K^\dag(x) & -\frac{3a(x)}{2\sqrt{6}}
	\end{pmatrix}\right) \, ,
\end{align}
where $\Sigma(x)$ contains all the pNGBs of the $SU(4)$ chiral symmetry, $\pi_8(x)\equiv \sum_{a=1}^8\pi^a(x)T_{SU(3)}^a$ is in the $SU(3)_c$ adjoint representation, $K(x)$ forms the $SU(3)_c$ anti-fundamental representation and $a(x)$ denotes the axion.
Singling out the axion direction, we have 
\begin{align}
    \Sigma_a(x) = \exp\begin{pmatrix}
        i\frac{a(x)}{ f_{\rm PQ}}\mathbf{1}_{3\times 3} & 0 \\[4pt]
		0 & -i\frac{3a(x)}{ f_{\rm PQ}}
    \end{pmatrix}\, ,
\end{align}
where $f_{\rm PQ}$ is defined as $f_{\rm PQ}\equiv \sqrt{6}F_a$ in analogy to the PQ breaking scale in the axion models with the PQ complex scalar. Their relation with the $SU(N)_V$ confinement scale $\Lambda_{\rm conf}$ is roughly given by ~\cite{Manohar:1983md,Marsh:2015xka}
\begin{align}
    f_{\rm PQ}/\sqrt{6} = F_{a} \sim \Lambda_{\rm conf}/(4\pi)\, .
\end{align}

\subsection{Matching conditions 
\label{sec:match}}

We assume that the $SU(4N)\times SU(3)_2\times SU(N)_2\times U(1)_X$ gauge symmetry is spontaneously broken into $SU(N)_V\times SU(3)_c\times U(1)_Y$ at a cut-off scale $\Lambda_{\rm cut}$, where fields $\Phi$, $\Phi'$, and $\Phi''$ obtain VEVs.
At the cut-off scale, the gauge couplings are related by
\begin{align}
  &  \frac{1}{g_{SU(4)_{1}}^2(\Lambda_{\rm cut})}= \frac{N}{g_{SU(4N)}^2(\Lambda_{\rm cut})}\ ,\label{eq:special_su4}\\[1ex]
  &  \frac{1}{g_{SU(N)_{1}}^2(\Lambda_{\rm cut})}=\frac{4}{g_{SU(4N)}^2(\Lambda_{\rm cut})} \ ,\label{eq:special_sun}
\end{align}
where $g_{SU(4N)}$ denotes the $SU(4N)$ gauge coupling, and the factors $N$ and $4$ in the right-hand sides originate from our special embedding of $SU(4)_1\subset SU(4N)$ and $SU(N)_1\subset SU(4N)$ (see Appendix~\ref{sec:special}).
The gauge couplings after the symmetry breaking are
\begin{align}
  &  \frac{1}{g_{c}^2(\Lambda_{\rm cut})}=\frac{1}{g_{SU(3)_2}^2(\Lambda_{\rm cut})}+\frac{1}{g_{SU(4)_1}^2(\Lambda_{\rm cut})} \ ,\\[1ex]
  &  \frac{1}{g_{Y}^2(\Lambda_{\rm cut})}=\frac{1}{g_{U(1)_X}^2(\Lambda_{\rm cut})}+\frac{24z^2}{g_{SU(4)_1}^2(\Lambda_{\rm cut})} \ ,\\[1ex]
  &  \frac{1}{g_{N}^2(\Lambda_{\rm cut})}=\frac{1}{g_{SU(N)_{2}}^2(\Lambda_{\rm cut})}+\frac{1}{g_{SU(N)_1}^2(\Lambda_{\rm cut})}\ .
\end{align} 
While the gauge couplings are not unified into a single parameter at the cut-off scale, they satisfy the above matching conditions,
and the scale $\Lambda_{\rm cut}$ is not determined because of the gauge couplings for $SU(3)_2\times SU(N)_2\times U(1)_X$.

Below the cut-off scale $\Lambda_{\rm cut}$ and above the $SU(N)_V$ confinement scale $\Lambda_{\rm conf}$,
the one-loop running of the gauge couplings can be expressed as
\begin{align}
  &  \frac{1}{\alpha_c(\mu)}=\frac{1}{\alpha_c (\Lambda_{\rm conf})}+\frac{1}{2\pi}
    \left(7-\frac{2N}{3} \right)\log(\mu/\Lambda_{\rm conf})\ ,\\[1ex]
  &  \frac{1}{\alpha_N(\mu)}=\frac{1}{\alpha_N(\Lambda_{\rm conf})}+\frac{1}{2\pi}
    \left(\frac{11}{3}N-\frac{8}{3} \right)\log(\mu/\Lambda_{\rm conf})\ ,
\end{align}
with $\alpha_c \equiv g_c^2/(4\pi)$ and $\alpha_N \equiv g_N^2/(4\pi)$.
At the confinement scale, the $SU(N)_V$ gauge coupling becomes strong, and we take
\begin{align}
    \frac{1}{\alpha_N(\Lambda_{\rm conf})}=0\ ,
\end{align}
which leads to
\begin{align}
    \frac{1}{\alpha_N(\Lambda_{\rm cut})}=\frac{1}{2\pi}  \left(\frac{11}{3}N-\frac{8}{3} \right)\log(\Lambda_{\rm cut}/\Lambda_{\rm conf})\ .
\end{align}
Combining the above relations, we obtain
\begin{align}
    \label{eq:coupling}
    \frac{1}{\alpha_c}
    &=\frac{1}{\alpha_{SU(3)_2}}+\frac{N}{\alpha_{4N}} \nonumber\\
    &=\frac{1}{\alpha_{SU(3)_2}}+\frac{N}{4}\frac{1}{\alpha_{SU(N)_1}} \nonumber\\
    &=\frac{N}{4}\frac{1}{\alpha_{N}}+\biggl(\frac{1}{\alpha_{SU(3)_2}}-\frac{N}{4}\frac{1}{\alpha_{SU(N)_2}}\biggr)
    \qquad\text{at}~\mu=\Lambda_{\rm cut}\ ,
\end{align}
with $\alpha_{4N} \equiv g_{SU(4N)}^2/(4\pi)$.
We can determine the value of $\alpha_{4N}$ for a given $\Lambda_{\rm conf},~\alpha_{SU(3)_2}^{-1},~N$, and $\Lambda_{\rm cut}$.

\begin{figure}
    \centering
    \includegraphics[width=0.325\linewidth]{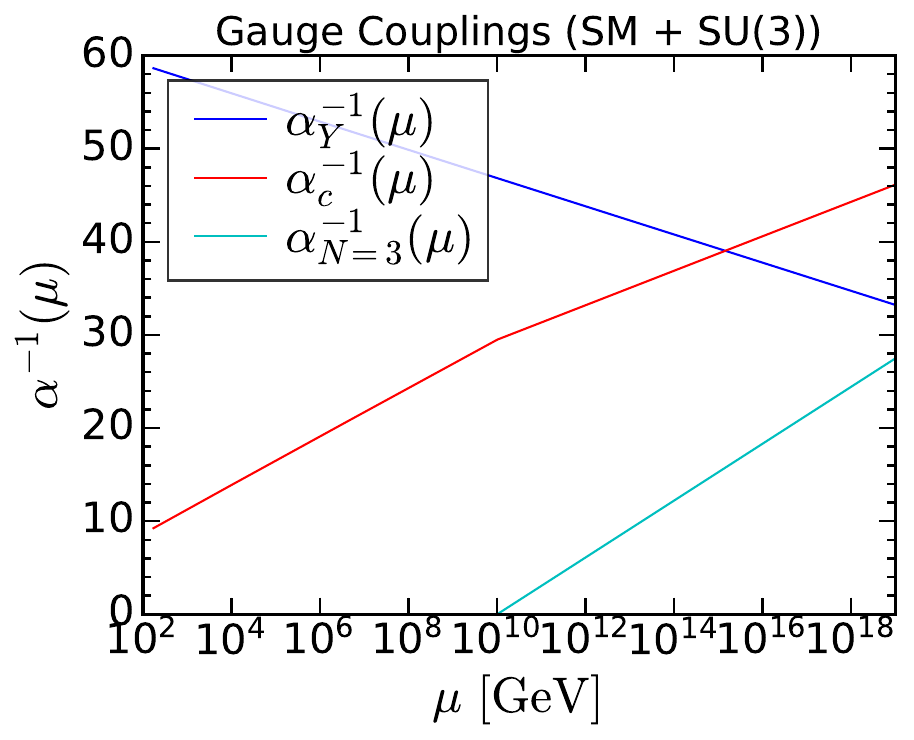}
    \includegraphics[width=0.325\linewidth]{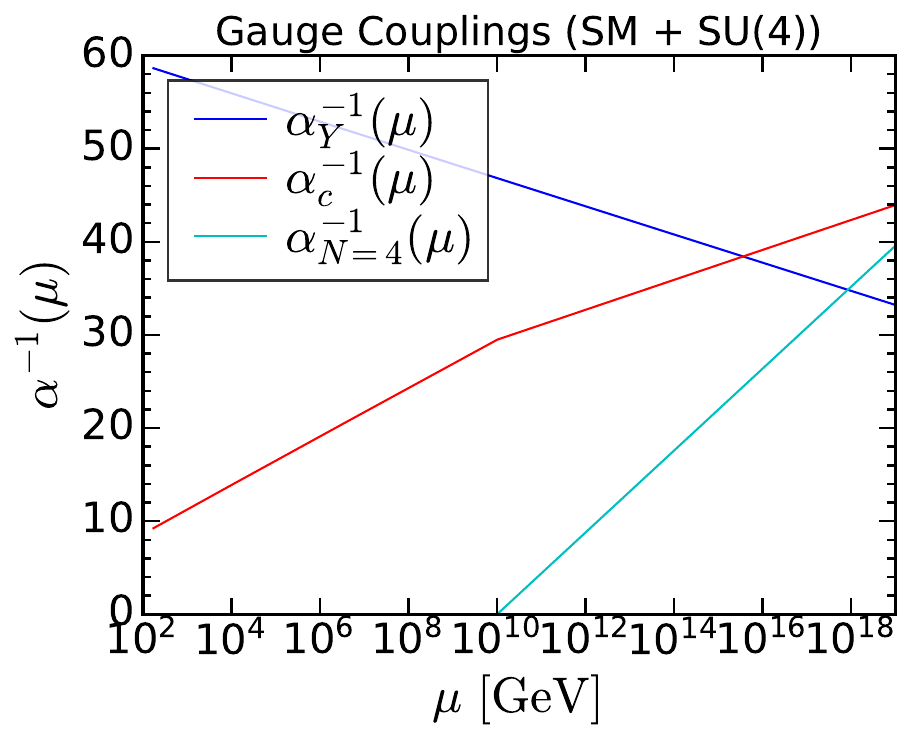}
    \includegraphics[width=0.325\linewidth]{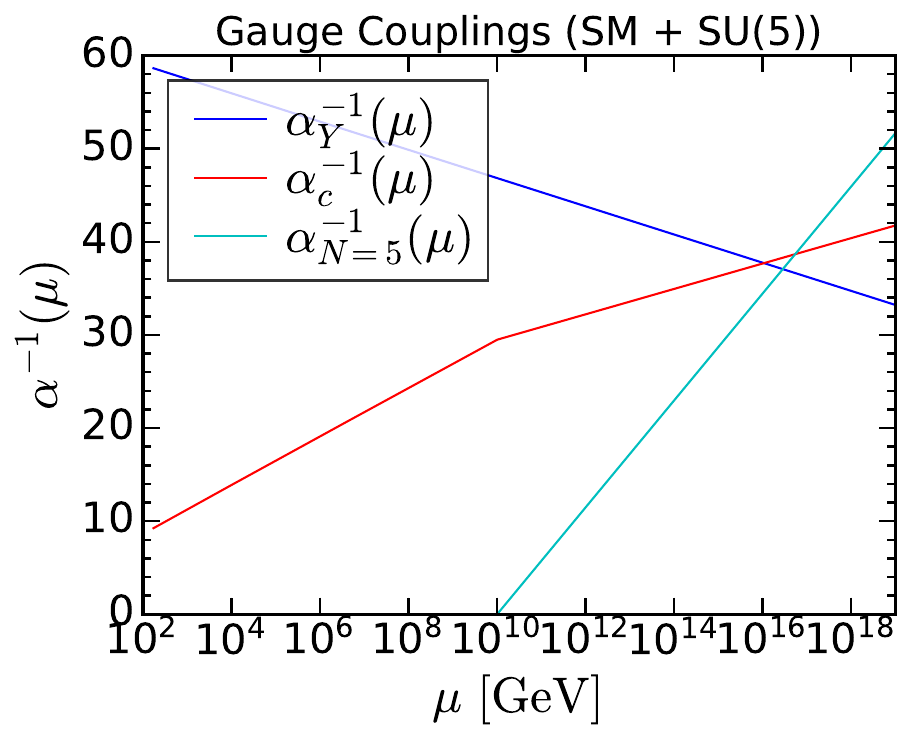}
    \caption{Renormalization group (RG) evolution of the gauge couplings for $N=3,4,5$. Here we take $\Lambda_{\rm conf}=10^{10}~{\rm GeV}$ as an example.
    }
    \label{fig:gaugecoupling}
\end{figure}

As an example, we can take $\Lambda_{\rm conf} =10^{10} \, \rm  GeV$, leading to 
\begin{align}
    \frac{1}{\alpha_c(\Lambda_{\rm conf} )} \approx 30 \ ,
\end{align}
as inferred from the SM $SU(3)_c$ gauge coupling flow, assuming all $SU(3)_c$ charged fields except for the SM quarks are around the confinement scale. Note that some mesons have masses slightly smaller than the confinement scale, but in good approximation, we add them to the gauge coupling beta function at $\Lambda_{\rm conf}$.
Figure~\ref{fig:gaugecoupling} shows the examples of the gauge coupling evolutions for $N=3,4,5$ below the cut-off scale in the case with $\Lambda_{\rm conf} =10^{10}~{\rm GeV}$.
The $U(1)_Y$ hypercharge gauge group is the linear combination of $U(1)_Z$ and $U(1)_X$. From the matching condition, it needs to satisfy $1/\alpha_Y > 24z^2N/\alpha_{4N}$ at $\mu =\Lambda_{\rm cut}$. For example, in the case with $N=4$, $a=1/6$, $b=-1/2$, and hence, we have $z=1/6$ and $x=0$, which requires $\alpha_Y^{-1} > 8/3 ~\alpha_{4N}^{-1}$ at $\mu =\Lambda_{\rm cut}$.

Next, we consider the matching conditions of $\theta$-terms between the UV and IR theories.
Their $\theta$-terms are respectively written as (see Appendix~\ref{sec:special})
\begin{align}
    {\rm UV: }& \quad \int \frac{\theta_{4N}g_{SU(4N)}^2}{8\pi^2}{\rm tr}(F_{4N}\wedge F_{4N}) +\frac{\theta_{3}g_{SU(3)_2}^2}{8\pi^2}{\rm tr}(F_{3}\wedge F_{3}) +\frac{\theta_{N_2}g_{SU(N)_2}^2}{8\pi^2}{\rm tr}(F_{N_2}\wedge F_{N_2}) ~,\nonumber\\[2ex]
    {\rm IR: }& \quad \int \frac{N\theta_{4N}g_{c}^2}{8\pi^2}{\rm tr}(F_{c}\wedge F_{c}) +\frac{\theta_{3}g_{c}^2}{8\pi^2}{\rm tr}(F_{c}\wedge F_{c}) \nonumber\\[1ex]
    &\quad +\int \frac{4\theta_{4N}g_{N}^2}{8\pi^2}{\rm tr}(F_{N}\wedge F_{N}) +\frac{\theta_{N_2}g_{N}^2}{8\pi^2}{\rm tr}(F_{N}\wedge F_{N}) ~,
\end{align}
where $F_{4N}$, $F_3$ and $F_{N_2}$ denote the gauge field strengths of $SU(4N)$, $SU(3)_2$ and $SU(N)_2$ respectively, while $F_c$ and $F_N$ are the $SU(3)_c$ and $SU(N)_V$ field strengths. 
The $\theta$-term matching conditions can, therefore, be depicted as
\begin{align}
    \theta_c = N\theta_{4N} +\theta_3~, \qquad \theta_N = 4\theta_{4N} +\theta_{N_2}~,
    \label{eq:theta}
\end{align}
which are the QCD $\theta$-angle and the $\theta$-angle for the $SU(N)_V$ gauge symmetry. Again, the factors $N$ and $4$ come from our special embedding.

The anomaly coefficients for $U(1)_A$ are given by
\begin{equation}
    {\cal A}\bigl( U(1)_A-SU(3)_c^2\bigr) =N ~,\qquad {\cal A}\bigl( U(1)_A-SU(N)_V^2\bigr) =4 ~,
\end{equation}
while those for $U(1)_{\rm PQ}$ are
\begin{equation}
    {\cal A}\bigl( U(1)_{\rm PQ}-SU(3)_c^2\bigr) =N ~,\qquad {\cal A}\bigl( U(1)_{\rm PQ}-SU(N)_V^2\bigr) =0~.
\end{equation}
After the $SU(N)_V$ confinement, $U(1)_A$ and $U(1)_{\rm PQ}$ symmetries are spontaneously broken, producing two pNGBs, $\theta_\eta$ and $\theta_a$, respectively.
Then, they couple to the $SU(3)_c$ and $SU(N)_V$ gauge fields with the forms,
\begin{align}
&\int (N\theta_a+N\theta_\eta+\theta_c)\frac{g_c^2}{8\pi^2} {\rm tr}(F_c\wedge F_c)~,\nonumber\\[1ex]
& \int (4\theta_\eta+\theta_N)\frac{g_N^2}{8\pi^2} {\rm tr}(F_N\wedge F_N)~.
\end{align}
Here, $\theta_a$ corresponds to the axion to solve the strong CP problem in this model. Similar to the axion, $\theta_\eta$ removes $\theta_N$ of the $SU(N)_V$ at the minimum of its potential generated by the $SU(N)_V$ non-perturbative dynamics. 
Note that $\theta_a$ and $\theta_\eta$ are the dimensionless phase component fields. They can be expressed as $\theta_a=a/f_{\rm PQ} $ and $\theta_\eta=\eta /f_{\rm PQ} $, where $a$ and $\eta$ are the normalized fields. 

Finally, the non-perturbative QCD effects generate the axion potential~\cite{GrillidiCortona:2015jxo},
\begin{align}
\label{QCDeffect}
    V_{\rm QCD}\approx
    -\frac{m_um_d}{(m_u+m_d)^2}m_\pi^2f_\pi^2 \cos\left(N\theta_a+\bar\theta_c\right)  ,
\end{align} 
where $\bar{\theta}_c\equiv \theta_c+N\langle\theta_\eta\rangle -{\rm arg}\,{\rm det}\left(M_uM_d\right)$,
$m_\pi$, $f_\pi$ denote the pion mass and decay constant, $m_u$ and $m_d$ are the up and down quark masses. The axion decay constant has the relation $f_a=f_{\rm PQ} /N$.
As we can see from this axion potential, the IR theory seems to have a domain wall number $N_{\rm DW}=N$, which leads to the domain wall problem. However, this can be resolved after including small instanton effects from the UV gauge theory, which will be explained in the next subsection.

\begin{figure}[t!]
    \centering
    \includegraphics[width=0.8\textwidth]{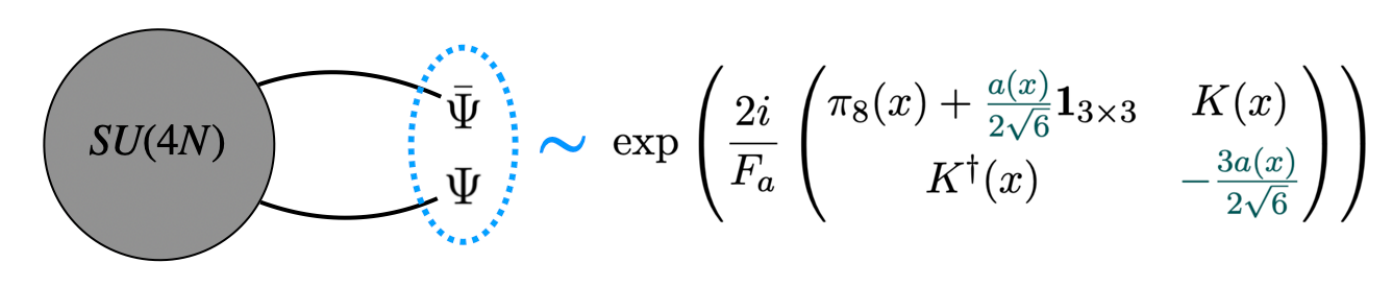} 
    \caption{'t Hooft vertex for the $SU(4N)$ instanton effect which generates the effective mass term for $\Psi,\bar{\Psi}$. The blue dashed circle denotes the fermion condensation under $SU(N)_V$ while the pNGB matrix on the right contains the axion. This interaction gives rise to the axion potential.}
    \label{fig:tHooft}  
\end{figure}

\subsection{Small instanton effects 
\label{sec:instanton}}

Assuming that the explicit $U(1)_{\rm PQ}$-breaking mass term of $\mathcal{L}\sim m\Psi\bar\Psi$ is absent,\footnote{The main goal of this work is to solve the domain wall problem in the composite axion model. The axion quality problem still remains in the current model.}
we can find the small instanton effect of the $SU(4N)$ gauge theory generating an effective quark mass term,
\begin{align}
    V_{I} = M_I \Psi \bar\Psi \sim C \Lambda_{\rm cut} \Psi\bar \Psi e^{-8\pi^2/(g_{SU(4N)}(\Lambda_{\rm cut}))^2}~,
\end{align}
where $C$ is a dimensionless constant which will be estimated below, $\Lambda_{\rm cut}$ denotes the cut-off scale where the $SU(4N)$ symmetry is spontaneously broken. 
Figure~\ref{fig:tHooft} illustrates the corresponding 't Hooft vertex.
According to the chiral perturbation theory, this mass term will lead to an axion potential given by 
\begin{align}
V_{\rm bias} = -\Lambda_{\rm conf}^3 M_I \tr\Sigma_a(x) + \mathrm{c.c.}= -\Lambda_{\rm conf}^3 M_I \lmk 3 e^{i\frac{a(x)}{f_{\rm PQ}}} + e^{-3i\frac{a(x)}{f_{\rm PQ}}}\rmk + \mathrm{c.c.}\, ,
\end{align}
Following the formalism of Instanton NDA~\cite{Csaki:2019vte,Csaki:2023ziz},%
\footnote{See~\cite{Sesma:2024tcd} for a functional method to evaluate instanton contributions to the axion potential in $SU(N)$ gauge theories with arbitrary matter content.}
the small instanton effect on the axion potential is estimated as
\begin{align}
    V_{\rm bias}&\approx - C_{4N}\biggl(\frac{2\pi}{\alpha_{4N}(\Lambda_{\rm cut})}\biggr)^{2\times 4N}\frac{1}{(4\pi)^2}\Lambda_{\rm cut}\Lambda_{\rm conf}^3\bigl(3e^{i\theta_a}+e^{-3i\theta_a}\bigr) e^{-4\pi^2}e^{-\frac{2\pi}{\alpha_{4N}(\Lambda_{\rm cut})}} + {\rm c.c.}\nonumber\\
    &= -2C_{4N}\biggl(\frac{2\pi}{\alpha_{4N}(\Lambda_{\rm cut})}\biggr)^{2\times 4N}\frac{1}{(4\pi)^2}\Lambda_{\rm cut}\Lambda_{\rm conf}^3 e^{-4\pi^2}e^{-\frac{2\pi}{\alpha_{4N}(\Lambda_{\rm cut})}}\bigl(3\cos{\theta_a}+\cos{3\theta_a}\bigr)~,
    \label{eq:small_instanton}
\end{align}
where $\Lambda_{\rm conf}$ denotes the confinement scale of $SU(N)_V$ and $\theta_a=\frac{a}{f_{\rm PQ}}$ denotes the dimensionless axion field. The instanton density is defined as~\cite{tHooft:1976snw,Bernard:1979qt,Csaki:2019vte,Csaki:2023ziz}
\begin{equation}
    C_{4N}=\frac{K_1e^{-( S^{(1/2)}-F^{(1/2)} ) \alpha(1/2)-( S^{(1)}-F^{(1)} ) \alpha(1)}}{(4N-1)!(4N-2)!}e^{-4NK_2}~,
\end{equation}
with $K_1\approx 0.466$, $K_2\approx 1.678$, $\alpha(1/2)\approx 0.145873$, $\alpha(1)\approx 0.443307$.
Here, $S^{(t)}$ and $F^{(t)}$ are the numbers of scalars and fermions, respectively, which transform in the isospin $t$ representation under the $SU(2)$ containing the instanton. In the present case, we have $S^{(1)}=8+(N^2-1)$, $S^{(1/2)}=(8+N^2-1)\times \bigl(2(4N-2)\bigr)+N$, $F^{(1)}=0$, $F^{(1/2)}=2$. 
This potential provides a bias term in the axion potential, and it corresponds to the domain wall number $N_{\rm DW}=1$. Combining $V_{\rm QCD}$ and $V_{\rm bias}$ leads to $N_{\rm DW}=1$, which can solve the domain wall problem.

\begin{table}
    \centering
    \renewcommand{\arraystretch}{1.5}
    \setlength{\tabcolsep}{1.5pt}
    \begin{tabular}{|c|c|c|c|c|c|c|c|c|c|}
        \hline
       \ $N$ \ & \ $SU(4N)$ \ & \ $1/\alpha_{4N}$ \ & \ $\Lambda_{\rm cut}~{\rm [GeV]}$ \ & \ $V_{{\rm bias},0}~{\rm [GeV^4]}$ \ \\ \hline
       $5$&    $SU(20)$ & $7.28 $ & $1.18\times10^{15} $ & $ 1.12\times 10^{-54}$ \\ \hline 
       $4$&    $SU(16)$ & $9.21 $ & $M_P$ & $6.3\times 10^{-20} $ \\ \hline
       $3$&    $SU(12)$ & $6.40 $ & $M_P$ & $2.1\times10^{4} $ \\ \hline
    \end{tabular}
    \vspace{3mm}
    \caption{The estimation of $SU(4N)$ small instanton effects with $\Lambda_{\rm conf}=4\pi F_a =10^{10}~{\rm GeV}$. First row: $1/{\alpha_{SU(3)_2}(\Lambda_{\rm cut})}=0$ and $1/{\alpha_{SU(N)_2}(\Lambda_{\rm cut})}=0$ are imposed. Second and third rows: $\Lambda_{\rm cut}=M_P$ and $1/{\alpha_{SU(N)_2}(\Lambda_{\rm cut})}=0$ are imposed, and $1/{\alpha_{SU(3)_2}(\Lambda_{\rm cut})}$ is determined by the matching condition. 
    }
    \label{tab:instanton}
\end{table}

We now estimate the size of small instanton effects. $V_{{\rm bias},0}$ and $V_{{\rm QCD},0}$ are defined to depict the size of the potentials as $V_{\rm bias}=-V_{{\rm bias},0}(\frac{3}{4} \cos{\theta_a}+\frac{1}{4}\cos{3\theta_a})$ and $V_{\rm QCD}=-V_{{\rm QCD},0}\cos{(N\theta_a+\bar{\theta}_c)}$~, so that they are positive and $\theta_a$ independent.
The gauge coupling $\alpha_{4N}$ at the cut-off scale $\Lambda_{\rm cut}$ can be obtained from Eq.~\eqref{eq:coupling}. We can consider different assumptions for the gauge couplings. First, we assume $\alpha_{SU(3)_2}^{-1}\sim \alpha_{SU(N)_2}^{-1} \sim 0$ at $\Lambda_{\rm cut}$ so that $1/\alpha_c=N/\alpha_{4N}=N/(4\alpha_N)$ at $\Lambda_{\rm cut}$. In this case, only when $N\geq 5$, there are solutions, which can be seen from Figure~\ref{fig:gaugecoupling}. For $N<5$, the gauge couplings $\alpha_c$ and $\alpha_N$ do not satisfy the relation before the Planck scale $M_P$. For $N=5$, the estimation of small instanton effects is shown in the first row of Table~\ref{tab:instanton}. Next, if we relax the constraint $1/{\alpha_{SU(3)_2}(\Lambda_{\rm cut})}= 0$, $\Lambda_{\rm cut}$ becomes a free parameter. Hence, we can consider the following scenario: $\Lambda_{\rm cut}=M_P=2.4\times 10^{18}~{\rm GeV}$, $1/{\alpha_{SU(N)_2}(\Lambda_{\rm cut})}=0$, and $1/{\alpha_{SU(3)_2}(\Lambda_{\rm cut})}$ is determined by the matching condition. The estimation of small instanton effects for $N=4$ and $N=3$ is shown in the second and third rows of Table~\ref{tab:instanton}, respectively.
Instanton NDA is a dimensional analysis, and it systematically accounts for the $4\pi$ factors appearing due to insertions of fermion zero modes or loop integrals. Its accuracy is generally expected to be limited to factors of a few, corresponding to ${\cal O}(1)$ uncertainties.

\begin{figure}
    \centering
    \includegraphics[width=0.49\linewidth]{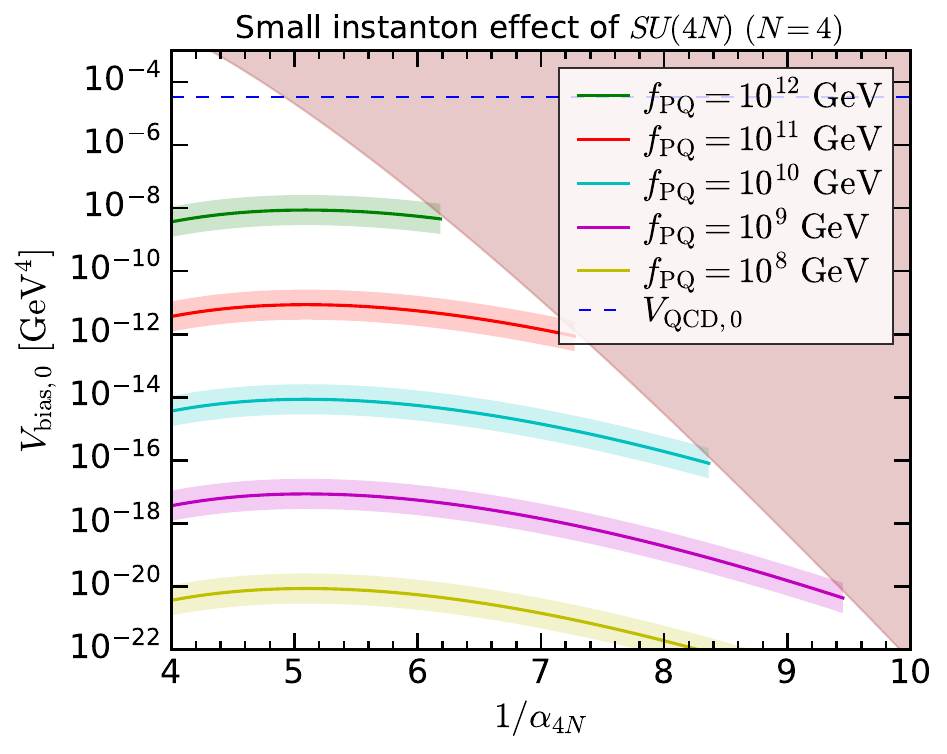}
    \includegraphics[width=0.49\linewidth]{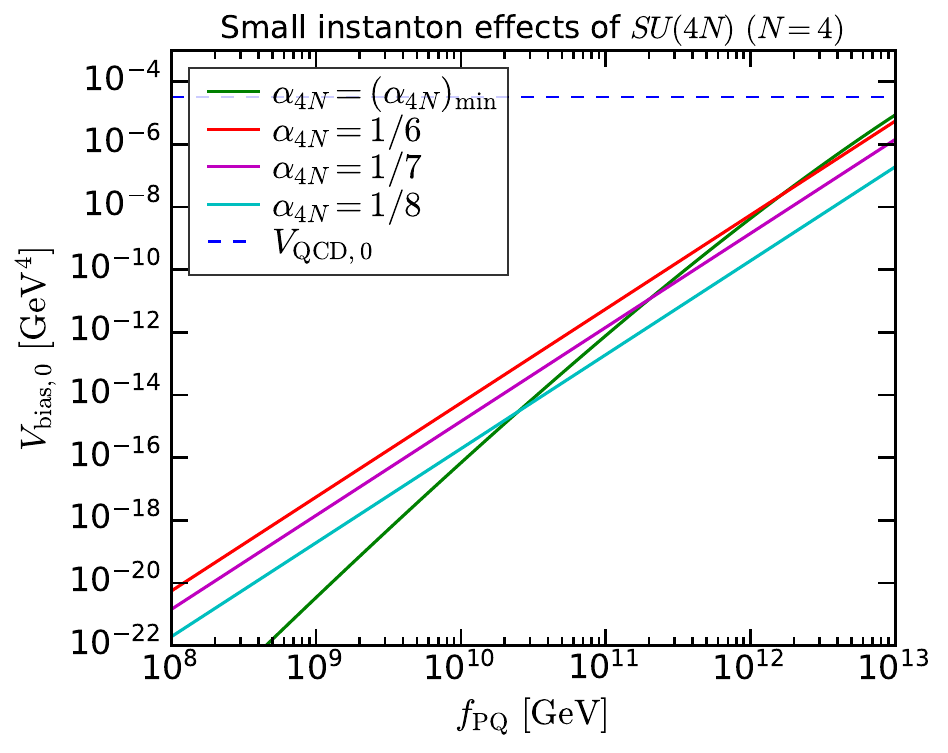}
    \caption{Small instanton effects on the axion potential for $N=4$. Solid lines are the estimated $V_{{\rm bias},0}$, while the dashed line is the axion potential generated from non-perturbative QCD effects. 
    The shaded bands around the solid lines depict the uncertainties of the NDA estimation with a factor of $3$.
    The pink-brown shaded region is excluded by the constraint $\alpha_{4N}<(\alpha_{4N})_{\rm min}$~.
    }
    \label{fig:Vbias}
\end{figure}

From Eq.~\eqref{eq:coupling}, $\alpha_{4N}$ can be treated as a free parameter with the constraint, $0\leq\alpha_{4N}^{-1}\leq {\rm min}\bigl(\frac{1}{N}\alpha_c^{-1},\, \frac{1}{4}\alpha_N^{-1}\bigr)$,
where $\Lambda_{\rm cut}=M_P$, and $1/{\alpha_{SU(3)_2}(\Lambda_{\rm cut})}$ and $1/{\alpha_{SU(N)_2}(\Lambda_{\rm cut})}$ are determined by the matching condition.
Small instanton effects of $SU(4N)$ are
suppressed by ${\rm exp}\bigl( -\frac{2\pi}{\alpha_{4N}}\bigr)$ as shown in Eq.~\eqref{eq:small_instanton}.
At the upper limit of $\alpha_{4N}^{-1}$, we have $\alpha_{4N} =4\alpha_N$ for $N=4$.
Figure~\ref{fig:Vbias} shows the small instanton effect on the axion potential $V_{\rm bias}$ for $N=4$ with respect to $\alpha_{4N}$ and $f_{\rm PQ}$. Solid lines are the estimated $V_{{\rm bias},0}$, while the dashed line is the axion potential generated from non-perturbative QCD effects. 
Here, $(\alpha_{4N})_{\rm min}=4\alpha_N $, and $6\lesssim (\alpha_{4N})_{\rm min}^{-1}\lesssim 10$ depends on $f_{\rm PQ}$ through RG evolution. When $\alpha_{4N}>1/4$ becomes strongly coupled, the perturbative computation starts to blow up, and hence the values of $V_{{\rm bias},0}$ become unreliable.
Again, we note that the plots shown in Figure~\ref{fig:Vbias} are subject to ${\cal O}(1)$ uncertainties arising from the NDA estimation.

\section{Axion cosmology
\label{sec:cosmology}}

Let us now explore cosmological implications of our composite axion model where
the axion potential is generated by non-perturbative QCD effects and small instanton effects, which can be described as
\begin{align}
    V(\theta_a)=V_{\rm QCD} +V_{\rm bias}~.
\end{align}
The two contributions $V_{\rm QCD}$ and $V_{\rm bias}$ dominate at different epochs in the cosmic history.
We consider the post-inflationary axion where the $U(1)_{\rm PQ}$ symmetry is spontaneously broken by the $SU(N)_V$ strong dynamics after reheating. The spontaneous breaking of the PQ symmetry gives rise to cosmic strings with the winding of $\theta_a:\, 0\to 2\pi$. 
Before non-perturbative QCD effects become effective, the axion starts to oscillate due to the axion mass $m_a$ generated by small instanton effects $V_{\rm bias}$, which can be depicted as
\begin{align}
    & m_a(T_1)\approx m_{\rm bias} 
    =3H(T_1)~,\\[1ex]
    & m_{\rm bias}^2=\frac{1}{f_{\rm PQ} ^2}\frac{\partial^2V_{\rm bias}}{\partial \theta_a^2}~,
\end{align}
where $T_1$ is the temperature when the oscillation starts, and $H(T_1)$ is the Hubble parameter. During the radiation-dominated period, the Hubble parameter can be expressed as
\begin{equation}
    H(T_1)^2 \approx \frac{\pi^2}{90 M_P^2}g_*T_1^4~,
\end{equation}
with $g_*$ the number of effective relativistic degrees of freedom of the energy density. 
We take 
$g_*\approx 70$ at $T\sim T_{1,\rm QCD}$ (see below) and 
$g_*\approx 106.75$ at $T\gtrsim 100~{\rm GeV}$. 
In our scenario, the axion starts to oscillate because of the small instanton effects instead of the QCD effects in the conventional case. Therefore, we require 
\begin{equation}
    T_1 \, > \, T_{1,{\rm QCD}}\,\equiv 0.98~{\rm GeV}\left( \frac{f_a}{10^{12}~{\rm GeV}} \right)^{-0.19}~,
\end{equation}
where $T_{1,{\rm QCD}}$ denotes the temperature when the axion would start to oscillate with non-perturbative QCD effects if there was no bias term $V_{\rm bias}$~\cite{Gross:1980br,Wantz:2009it,Wantz:2009mi}.
Since the axion potential generated from small instanton effects $V_{\rm bias}$ corresponds to domain wall number $N_{\rm DW}=1$, the domain wall problem is addressed.
Below the temperature $\sim T_1$, a cosmic string is attached by a domain wall, but the domain wall tension can collapse the string-wall system. 
In approximation, we assume the collapse time to be around $T_1$. A more precise collapse time can be determined by a dedicated simulation, which lies beyond the scope of the present work. 

\begin{figure}
    \centering
    \includegraphics[width=0.49\linewidth]{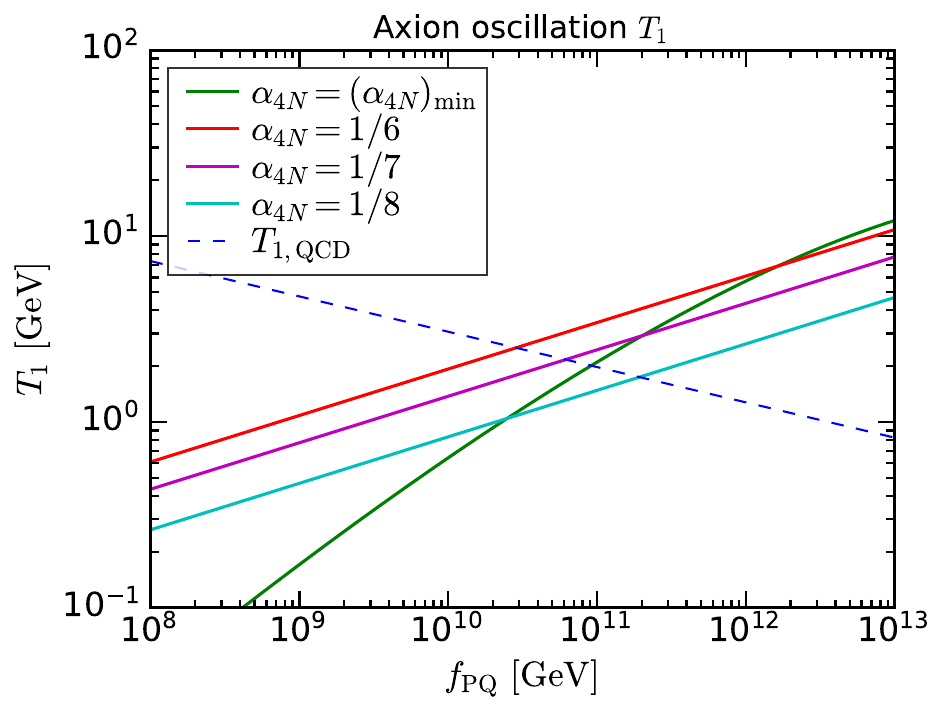}
    \includegraphics[width=0.49\linewidth]{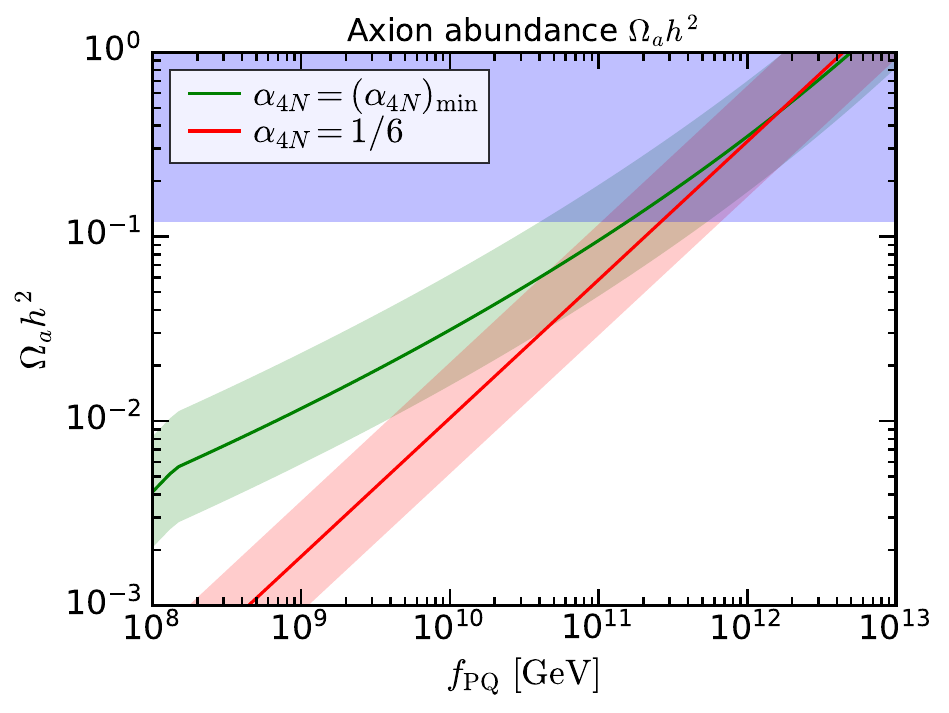}
    \caption{The cosmological implications of our model and constraints on the parameter space. The left panel describes the temperature $T_1$ where the axion starts to oscillate due to the mass induced by small instanton effects, and the dashed line is $T_{1,{\rm QCD}}$ in the conventional case. The right panel shows the axion abundance as the solid lines, with shaded bands indicating the uncertainties. The blue shaded region is excluded by the DM abundance constraint.}
    \label{fig:T1_DM}
\end{figure}

The axion abundance in our scenario is estimated in the similar way as the case of $N_{\rm DW}=1$, and it can be depicted as~\cite{Lyth:1991bb,Hiramatsu:2012gg,Kawasaki:2014sqa}
\begin{align}
    \Omega_a h^2 &\approx (2^{+2}_{-1})\times 10^{-12}\, \frac{f_{\rm PQ}}{T_1}~.
    \label{eq:DM}
\end{align}
We note that this is an order-of-magnitude estimate, and the quoted range is intended to reflect this level of approximation.%
\footnote{
The estimate assumes that the domain walls collapse around $T_1$. A more reliable determination of the precise collapse time requires dedicated numerical simulations.
}
Note that this is an initial estimation of the axion abundance, and a precise abundance remains uncertain and may differ from this estimation. Considering the axion as (a part of) dark matter (DM), the constraint from the DM relic abundance is given as $\Omega_a h^2 < 0.12$~.
The lower bound for the axion decay constant is $f_a \gtrsim 4\times 10^8~{\rm GeV}$ obtained from the stellar cooling neutrino emission of SN 1987A~\cite{Carenza:2019pxu}.

Below the QCD scale, the axion potential is dominated by non-perturbative QCD effects, and the axion mass generated from $V_{\rm QCD}$ is 
\begin{equation}
    m_{a,{\rm QCD}}^2=\frac{1}{f_{\rm PQ} ^2}\frac{\partial^2V_{\rm QCD}}{\partial\theta_a^2}=\frac{m_um_d}{(m_u+m_d)^2}\frac{m_\pi^2f_\pi^2 }{f_a^2}~.
\end{equation}
The total axion mass is $m_a^2=m_{a,{\rm QCD}}^2+m_{\rm bias}^2$~, with $m_{\rm bias}\ll m_{a,{\rm QCD}}$.

The small instanton bias term is introduced in order to solve the domain wall problem. However, a bias term will shift the potential minimum of the QCD axion potential originated from $V_{\rm QCD}$, which may spoil the axion solution to the strong CP problem. The shift of the potential is estimated as
\begin{equation}
    \Delta \bar{\theta}_c\approx \frac{V_{{\rm bias},0}}{V_{{\rm QCD},0}}\bar{\theta}_c~.
\end{equation}
We require
\begin{equation}
    \Delta \bar{\theta}_c\lesssim 10^{-10}~,
\end{equation}
to satisfy the experimental upper bound of the neutron EDM~\cite{Abel:2020pzs} so that the axion field can still resolve the strong CP problem.

\begin{figure}
    \includegraphics[width=0.49\linewidth]{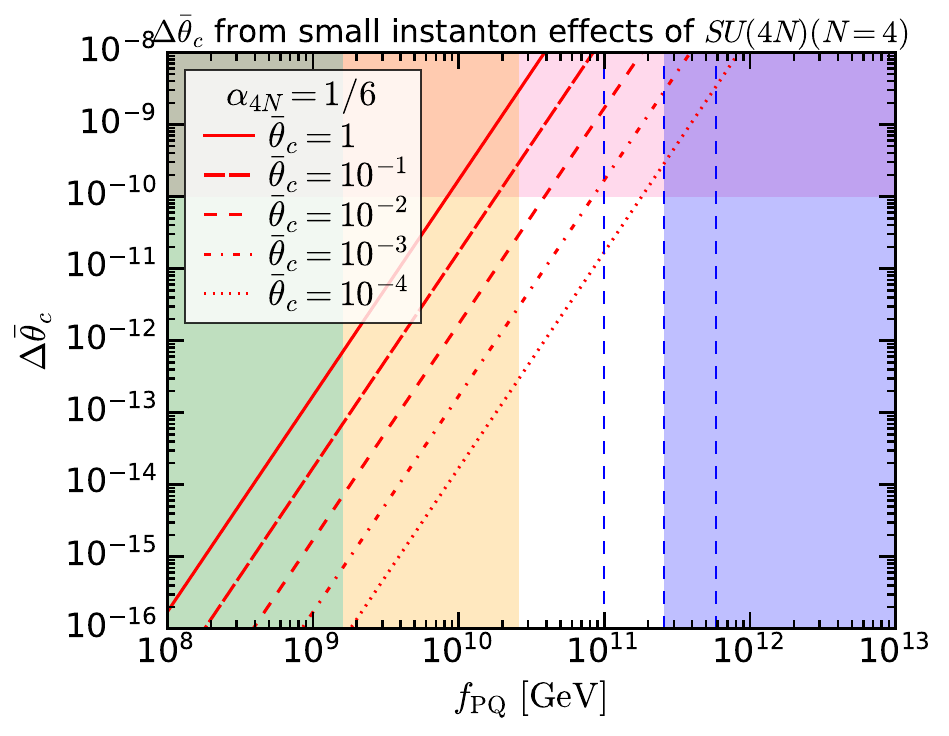}
    \includegraphics[width=0.49\linewidth]{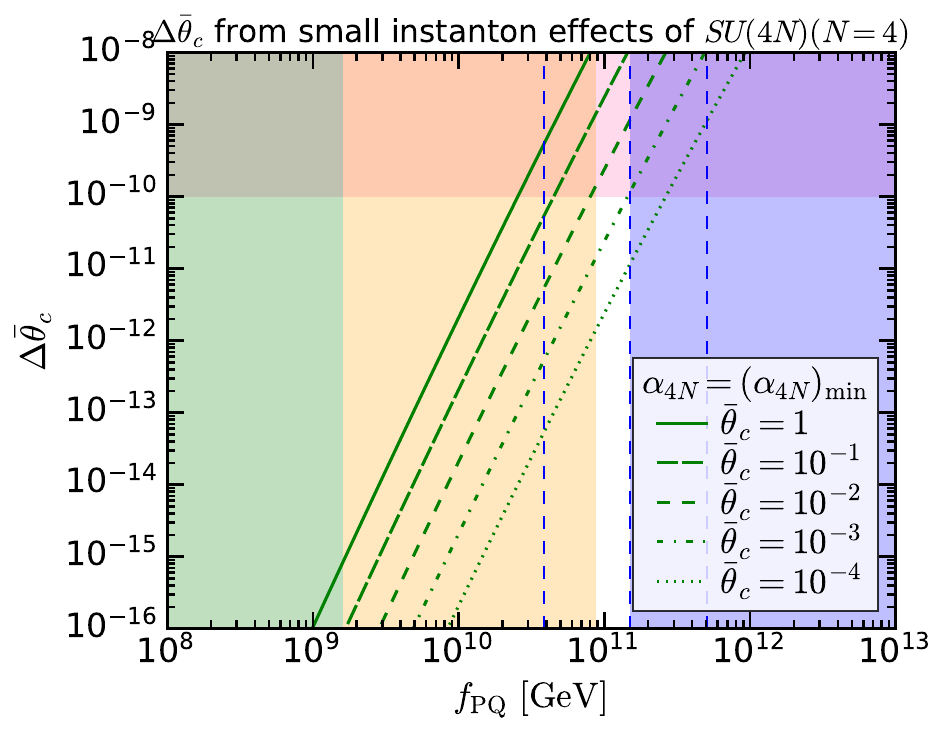}
    \includegraphics[width=0.75\linewidth]{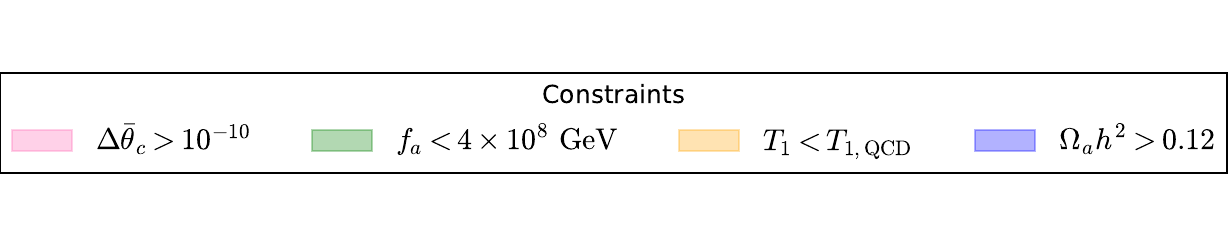}
    \vspace{-3mm}
    \caption{The shift $\Delta\bar{\theta}_c$ coming from small instanton effects of $SU(4N)$ with $N=4$. The shaded regions are the constraints on $f_{\rm PQ}$ and $\Delta\bar{\theta}_c$. The blue dashed lines denote the uncertainties of the DM abundance constraint in our estimation.
    }
    \label{fig:constraint}
\end{figure}

\begin{figure}
    \includegraphics[width=0.5\linewidth]{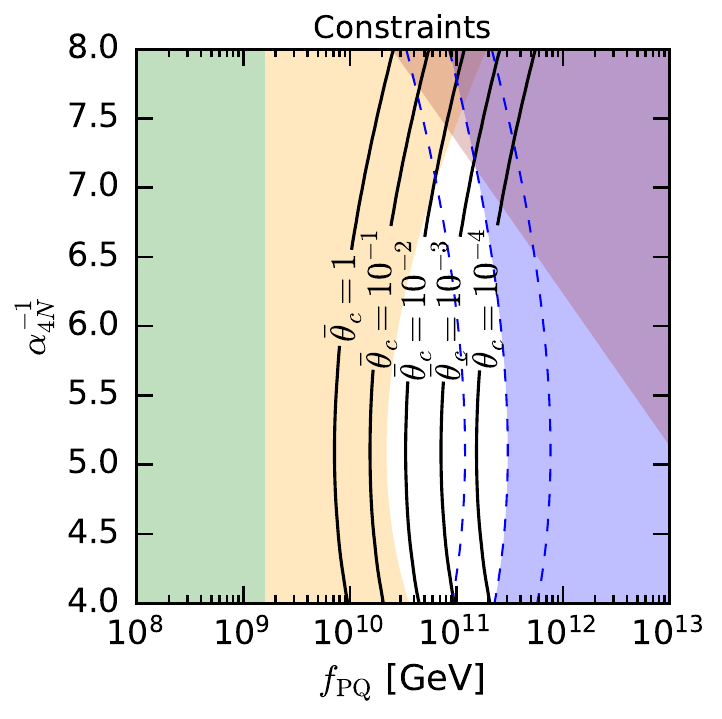}
    \includegraphics[width=0.75\linewidth]{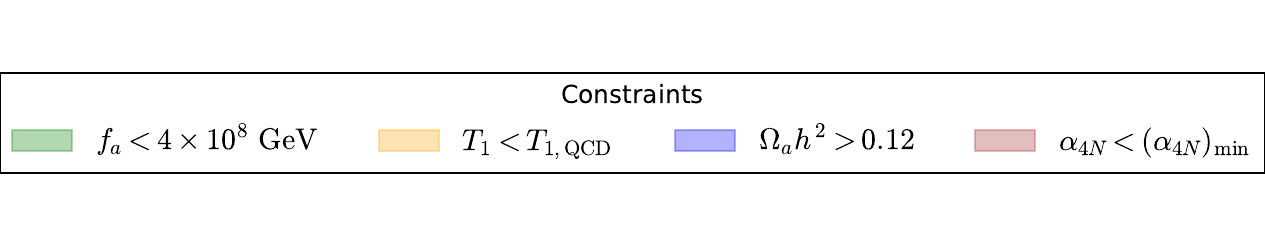}
    \caption{Constraints on the $f_{\rm PQ}-\alpha_{4N}^{-1}$ plane. The blue dashed lines represent the uncertainties of the DM abundance constraint in our estimation. The black solid lines indicate contours of $\Delta\bar{\theta}_c=10^{-10}$ for different values of $\bar{\theta}_c$. 
    }
    \label{fig:constraint2}
\end{figure}

Figure~\ref{fig:T1_DM} shows the cosmological implications of our model. In the left panel, solid lines represent the temperature $T_1$ where the axion starts to oscillate due to the mass induced by small instanton effects, while the dashed line is $T_{1,{\rm QCD}}$ in the conventional case. The focused parameter space is above the dashed line. The right panel shows the axion abundance as solid lines,\footnote{In the right panel of Figure~\ref{fig:T1_DM}, the kink in the green line at $f_{\rm PQ}\approx 10^8~{\rm GeV}$ arises from the transition in $(\alpha_{4N})_{\rm min}={\rm max}\bigl(N\alpha_c,\, 4\alpha_N\bigr)$~.} with shaded bands indicating the uncertainties. The blue shaded region represents the DM abundance constraint.

Figure~\ref{fig:constraint} describes the shift $\Delta\bar{\theta}_c$ coming from small instanton effects of $SU(4N)$ with $N=4$. The shaded regions are the constraints on $f_{\rm PQ}$ and $\Delta\bar{\theta}_c$. The orange-shaded, green-shaded, pink-shaded and blue-shaded regions represent $T_1<T_{1,{\rm QCD}}$, $f_a<4\times 10^{8}~{\rm GeV}$, $\Delta\bar{\theta}_c > 10^{-10}$ and $\Omega_ah^2>0.12$, respectively.
Finally, the constraints are summarized in the $f_{\rm PQ}-\alpha_{4N}^{-1}$ plane in Figure~\ref{fig:constraint2}. The shaded regions are ruled out by different constraints. The blue dashed lines are the uncertainties of the DM abundance constraint in our estimation. The black solid lines indicate contours of $\Delta\bar{\theta}_c=10^{-10}$ for different values of $\bar{\theta}_c$.
The white region denotes the viable parameter space. Our model requires $f_{\rm PQ}\gtrsim 10^{10}~{\rm GeV}$ and $\bar{\theta}_c\lesssim 10^{-1}$.
In the viable region, the axion abundance $\Omega_ah^2$ is a slightly below the observed DM relic density. To account for the full dark matter abundance, $\bar{\theta}_c \lesssim 10^{-3}$ is required.
We note that $T_1$ and $\Omega_ah^2$ are derived from $V_{\rm bias}$, and therefore we expect the corresponding constraints in Figure~\ref{fig:constraint} and Figure~\ref{fig:constraint2} to have ${\cal O}(1)$ uncertainties.

In order to achieve a small $\bar{\theta}_c$ (such as $\sim 10^{-4}$ in Figure~\ref{fig:constraint2}), we can consider spontaneous CP violation to eliminate the $\theta$-parameters for $SU(4N) \times SU(3)_2\times SU(N)_2$ at the tree-level and the Nelson-Barr (NB) mechanism~\cite{Nelson:1983zb,Barr:1984qx,Barr:1984fh} in the $SU(3)_2\times U(1)_X$ sector
to transfer CP violation into the Cabibbo-Kobayashi-Maskawa phase. We can introduce a pair of fermions $\psi=(\mathbf{1},\mathbf{3},\mathbf{1},-1/3)$ and $\bar{\psi}=(\mathbf{1},\bar{\mathbf{3}},\mathbf{1},1/3)$, and interactions, 
\begin{align}
    {\cal L}_{\rm NB}\sim \mu\psi\bar{\psi}+\sum_{f=1,2,3,\alpha=1,2} a_{\alpha f}\eta_{\alpha} \psi\bar{d}_f+...~,
\end{align}
where $\bar{d}_f$ represent the SM down-type quarks, $\mu$ is a mass parameter and $a_{\alpha f}$ are dimensionless coefficients. The complex scalar fields $\eta_\alpha$ acquire VEVs and spontaneously break CP symmetry.%
\footnote{When the VEVs of the scalar fields are far below the Planck scale, this also causes a hierarchy problem. However, the problem is less severe compared to the standard NB mechanism, since our setup does not aim to solve the strong CP problem.}
While this setup makes it possible to achieve $\bar{\theta}_c \ll {\cal O}(1)$,
we cannot expect that the NB mechanism itself solves the strong CP problem due to radiative corrections
\cite{Dine:2015jga,Fujikura:2022sot,Girmohanta:2022giy}.

\section{Conclusions and discussion
\label{sec:Discussion}}

We have presented a novel framework in which the cosmological domain wall problem of a post-inflationary composite axion model is resolved through the special embedding of the model into a UV theory with a larger gauge group. In conventional composite axion constructions, the color anomaly typically implies a domain wall number greater than unity, leading to stable string-wall networks that overclose the Universe. Our key observation is that the IR confining gauge group responsible for the composite axion, together with QCD, can be embedded into a larger UV gauge structure in a non-trivial way. In this setup, the UV theory essentially has domain wall number one, while the apparent vacuum degeneracy of the IR description is lifted by small instanton effects associated with the UV gauge dynamics. As a result, the domain walls become unstable and decay sufficiently early, while the PQ mechanism remains operative and the strong CP problem is still solved.

To demonstrate the idea explicitly, we constructed a concrete model, estimated the induced bias term in the axion potential from small instantons and identified parameter regions where it is large enough to remove the dangerous domain walls, but sufficiently small to avoid reintroducing an unacceptable shift of the QCD $\theta$-parameter. This study provides a controlled realization of a composite axion model in which the domain wall problem is cosmologically harmless.
Since the small instanton contribution can trigger axion oscillations earlier than in the conventional post-inflationary case, the resulting axion abundance is typically reduced. We found a viable region satisfying astrophysical constraints, relic abundance bounds, and the neutron EDM limit. In the minimal setup, however, the axion abundance tends to be smaller than the observed DM density, suggesting either that the axion constitutes only a sub-component of DM or that the additional structure of spontaneous CP violation operates.

Another important aspect of the original composite axion framework is the presence of exotic hadronic states arising from the new confining dynamics. The low-energy spectrum generically contains colored mesons and baryons that can be stable or long-lived in the absence of additional interactions. Such relics would lead to severe cosmological difficulties, since strongly interacting heavy particles may survive thermal freeze-out and subsequently form charged bound states after the QCD phase transition. We therefore considered higher-dimensional operators that connect the exotic sector to SM fields and induce sufficiently rapid decays of these states. Since these interactions must typically be suppressed by a cutoff scale not far above the confinement scale in order to ensure prompt enough decays, it is natural to ask whether the SM quarks participating in such interactions may themselves originate from related strong dynamics or partial compositeness. This possibility would provide a more coherent ultraviolet interpretation of the required operators and could relate the SM flavor structure to the composite axion sector. Exploring frameworks in which the relevant SM quarks and the exotic states share a common composite origin would therefore be an especially interesting direction for a future study.

A further direction concerns the axion quality problem
\cite{Dine:1986bg,Barr:1992qq,Kamionkowski:1992mf,Kamionkowski:1992ax,Holman:1992us,Kallosh:1995hi,Carpenter:2009zs}, which is not resolved in the original composite axion model that we specially embedded in the present work. Realistic axion model building ultimately requires a sufficient protection of the PQ symmetry against unwanted explicit breaking, especially from Planck-suppressed operators. It would therefore be highly interesting to apply the idea of special embedding to composite axion models that are specifically designed to possess high-quality PQ symmetries. In such a framework, the same ultraviolet group-theoretic structure may simultaneously control the domain wall number, generate a cosmologically viable bias term, and suppress dangerous PQ-violating operators. Exploring whether special embedding can unify these ingredients into a fully consistent and natural composite axion scenario would be an important subject for future investigation.

\section*{Acknowledgments}

Y.N. is supported by Natural Science Foundation of Shanghai. M.S. is supported by the MUR projects 2017L5W2PT.
M.S. also acknowledges the European Union - NextGenerationEU, in the framework of the PRIN Project “Charting unexplored avenues in Dark Matter” (20224JR28W).


\appendix

\section{Special embedding}\label{sec:special}

We follow the idea of special embedding developed in Refs.~\cite{Yamatsu:2015npn,Yamatsu:2017mei,Hor:2025gxo}, and
consider the symmetry breaking pattern,
\begin{align}
SU(MN)\ \to\ SU(M)\times SU(N)\,.
\end{align}
Our goal here is to embed the bifundamental representation of $SU(M)\times SU(N)$ into the fundamental representation of $SU(MN)$.
In $SU(M)\times SU(N)$, we represent the bifundamental as a single column vector $\mathbf{v}$, which transforms as
\begin{align}
\mathbf{v}\ \to\ e^{i \beta_M^a \hat T_M^a}\, e^{i \beta_N^\alpha \hat \tau_N^\alpha}\,\mathbf{v}\,,
\end{align}
where $\beta_M^a$ and $\beta_N^\alpha$ are real parameters. The generators $\hat T_M^a$ are defined by
\begin{align}
\hat T_M^a \equiv T_M^a \otimes \mathbbm{1}_{N\times N}
= \begin{pmatrix}
T_M^a & \mathbf{0} & \cdots \\
\mathbf{0} & T_M^a & \cdots \\
\vdots & \vdots & \ddots
\end{pmatrix},
\end{align}
where $T_M^a$ are the generators of $SU(M)$ in the fundamental representation, given by $M\times M$ Hermitian traceless matrices. The matrix $\mathbbm{1}_{N\times N}$ denotes the $N\times N$ identity matrix, and we normalize the generators as ${\rm tr}(T_M^a T_M^b)=\frac{1}{2}\delta^{ab}$.
Here, we define the tensor product as
\(
    A\otimes B=
    \begin{pmatrix}
        A B_{11} & A B_{12} & ...\\
          A B_{21} & A B_{22} & ...\\
          \vdots & \vdots 
    \end{pmatrix}
\)
where $A,~B$ denote matrices.
Similarly, the generators $\hat \tau_N^\alpha$ are defined as
\begin{align}
\hat \tau_N^\alpha \equiv \mathbbm{1}_{M\times M}\otimes \tau_N^\alpha
=
\begin{pmatrix}
\mathbbm{1}_{M\times M} (\tau_N^\alpha)_{11} & \mathbbm{1}_{M\times M} (\tau_N^\alpha)_{12} & \cdots \\
\mathbbm{1}_{M\times M} (\tau_N^\alpha)_{21} & \mathbbm{1}_{M\times M} (\tau_N^\alpha)_{22}& \cdots \\
\vdots & \vdots & \ddots
\end{pmatrix},
\end{align}
where $\tau_N^\alpha$ are the generators of $SU(N)$ in the fundamental representation, given by $N\times N$ Hermitian traceless matrices.
One can verify that both $\hat T_M^a$ and $\hat \tau_N^\alpha$ are traceless Hermitian matrices, and that they are orthogonal:
\begin{align}
{\rm tr}(\hat T_M^a \hat \tau_N^\alpha)=0\,.
\end{align}
We now embed these generators into those of $SU(MN)$ as
\begin{align}
\hat T_M^a = \mathcal{O}_M^{am} T_{\rm UV}^m \ ,\qquad
\hat \tau_N^\alpha = \mathcal{O}_N^{\alpha m} T_{\rm UV}^m\,,
\end{align}
where $T_{\rm UV}^m$ are the $SU(MN)$ generator of $NM\times NM$  Hermitian traceless  matrices, normalized as
\begin{align}
{\rm tr}(T_{\rm UV}^m T_{\rm UV}^n)=\frac{1}{2}\delta^{mn}\,.
\end{align}
Using the properties of the tensor product, we obtain
\begin{align}
{\rm tr}(\hat T_M^a \hat T_M^b) &= N \cdot \frac{1}{2}\delta^{ab} \ ,\\[1ex]
{\rm tr}(\hat \tau_N^\alpha \hat \tau_N^\beta) &= M \cdot \frac{1}{2}\delta^{\alpha\beta}\,,
\end{align}
which imply
\begin{align}
\mathcal{O}_M^{am}\mathcal{O}_M^{bn}\delta_{mn} &= c_M\,\delta^{ab} \ , \qquad c_M = N \, ,\\[1ex]
\mathcal{O}_N^{\alpha m}\mathcal{O}_N^{\beta n}\delta_{mn} &= c_N \, \delta^{\alpha\beta} \ , \qquad c_N = M\,.
\end{align}

Let us consider a fermion $\psi$ which transforms as the fundamental representation of $SU(MN)$ and the bifundamental representation of $SU(M)\times SU(N) \subset SU(MN)$.
The covariant derivative is given by
\begin{align}
\mathcal{D}_\mu \psi
&= \partial_\mu \psi - i g_{\rm UV} A^m_{{\rm UV},\mu} T_{\rm UV}^m \psi \\[1ex]
&\supset \partial_\mu \psi
- i g_{{\rm IR},M} A^a_{{\rm IR},M,\mu} \hat T_M^a \psi
- i g_{{\rm IR},N} A^\alpha_{{\rm IR},N,\mu} \hat \tau_N^\alpha \psi\, , 
\end{align}
where $A^m_{{\rm UV},\mu}$ denotes the $SU(MN)$ gauge field with gauge coupling $g_{\rm UV}$,
while $A^a_{{\rm IR},M,\mu}$ and $A^\alpha_{{\rm IR},N,\mu}$ are the $SU(M)$ and $SU(N)$ gauge fields with gauge couplings
$g_{{\rm IR},M}$ and $g_{{\rm IR},N}$, respectively. 
Matching the gauge fields, we find
\begin{align}
A^m_{{\rm UV},\mu} = \frac{g_{{\rm IR},M}}{g_{\rm UV}} A^a_{{\rm IR},M,\mu}\, \mathcal{O}_M^{am}, \qquad 
A^m_{{\rm UV},\mu} = \frac{g_{{\rm IR},N}}{g_{\rm UV}} A^\alpha_{{\rm IR},N,\mu}\, \mathcal{O}_N^{\alpha m}\,.
\end{align}
Requiring canonical normalization of the gauge kinetic terms leads to
\begin{align}
g_{{\rm IR},M} = \frac{g_{\rm UV}}{\sqrt{c_M}} \ ,\qquad
g_{{\rm IR},N} = \frac{g_{\rm UV}}{\sqrt{c_N}} \ .
\end{align}
Finally, the topological term for the $SU(MN)$ gauge field is decomposed as
\begin{align}
\int \frac{g_{\rm UV}^2}{8\pi^2} {\rm tr}(F_{\rm UV}\wedge F_{\rm UV})
\ \supset\ 
\int c_N \frac{g_{{\rm IR},N}^2}{8\pi^2} {\rm tr}(F_{{\rm IR},N}\wedge F_{{\rm IR},N})
+ \int c_M \frac{g_{{\rm IR},M}^2}{8\pi^2} {\rm tr}(F_{{\rm IR},M}\wedge F_{{\rm IR},M})\,,
\end{align}
where $F_{\rm UV}$, $F_{{\rm IR},M}$ and $F_{{\rm IR},N}$ represent the $SU(MN)$, $SU(M)$ and $SU(N)$ gauge field strengths, respectively.

In the present work for the composite axion model, we focus on the special embedding of $SU(4N)\to SU(4) \times SU(N)$. Table~\ref{tab:su12}, \ref{tab:su16}, and \ref{tab:su20} summarize the branching rules of $su(4N)\supset su(4)\oplus su(N)$ for $N=3,4,5$, based on Ref.~\cite{Yamatsu:2015npn}.

\begin{table}[t!]
    \centering
    \renewcommand{\arraystretch}{1}
    \begin{tabular}{rl}
        \hline
         $su(12)$  & $su(4)\oplus su(3)$\\
         \hline
         $\mathbf{1}$ & $(\mathbf{1},\, \mathbf{1})$ \\
         $\mathbf{12}$ & $(\mathbf{4},\, \mathbf{3})$\\
         $\overline{\mathbf{12}}$ & $(\overline{\mathbf{4}},\, \overline{\mathbf{3}})$\\
         $\mathbf{66}$ & $(\mathbf{10},\, \overline{\mathbf{3}}) \oplus (\mathbf{6},\, \overline{\mathbf{6}})$\\
         $\overline{\mathbf{66}}$ & $(\overline{\mathbf{10}},\, \mathbf{3}) \oplus (\mathbf{6},\, \mathbf{6})$\\
         $\mathbf{78}$ & $(\mathbf{10},\, \overline{\mathbf{6}}) \oplus (\mathbf{6},\, \overline{\mathbf{3}})$\\
         $\overline{\mathbf{78}}$ & $(\overline{\mathbf{10}},\, \mathbf{6}) \oplus (\mathbf{6},\, \mathbf{3})$\\
         $\mathbf{143}$ & $(\mathbf{15},\, \mathbf{8})\oplus (\mathbf{15},\, \mathbf{1})\oplus (\mathbf{1},\, \mathbf{8})$\\
       \hline
    \end{tabular}
    \caption{Branching rules of special embedding $su(12)\supset su(4)\oplus su(3)$~\cite{Yamatsu:2015npn}.
    \vspace{5mm}}
    \label{tab:su12}
\end{table}

\begin{table}[h!]
    \centering
    \renewcommand{\arraystretch}{1}
    \begin{tabular}{rl}
        \hline
         $su(16)$  & $su(4)\oplus su(4)$\\
         \hline
         $\mathbf{1}$ & $(\mathbf{1},\, \mathbf{1})$ \\
         $\mathbf{16}$ & $(\mathbf{4},\, \mathbf{4})$\\
         $\overline{\mathbf{16}}$ & $(\overline{\mathbf{4}},\, \overline{\mathbf{4}})$\\
         $\mathbf{120}$ & $(\mathbf{10},\, \mathbf{6}) \oplus (\mathbf{6},\, \mathbf{10})$\\
         $\overline{\mathbf{120}}$ & $(\overline{\mathbf{10}},\, \mathbf{6}) \oplus (\mathbf{6},\, \overline{\mathbf{10}})$\\
         $\mathbf{136}$ & $(\mathbf{10},\, \mathbf{10}) \oplus (\mathbf{6},\, \mathbf{6})$\\
         $\overline{\mathbf{136}}$ & $(\overline{\mathbf{10}},\, \overline{\mathbf{10}}) \oplus (\mathbf{6},\, \mathbf{6})$\\
         $\mathbf{255}$ & $(\mathbf{15},\, \mathbf{15})\oplus (\mathbf{15},\, \mathbf{1})\oplus (\mathbf{1},\, \mathbf{15})$\\
       \hline
    \end{tabular}
    \caption{Branching rules of special embedding $su(16)\supset su(4)\oplus su(4)$~\cite{Yamatsu:2015npn}.
    \vspace{5mm}}
    \label{tab:su16}
\end{table}

\begin{table}[h!]
    \centering
    \renewcommand{\arraystretch}{1}
    \begin{tabular}{rl}
        \hline
         $su(20)$  & $su(4)\oplus su(5)$\\
         \hline
         $\mathbf{1}$ & $(\mathbf{1},\, \mathbf{1})$ \\
         $\mathbf{20}$ & $(\mathbf{4},\, \mathbf{5})$\\
         $\overline{\mathbf{20}}$ & $(\overline{\mathbf{4}},\, \overline{\mathbf{5}})$\\
         $\mathbf{190}$ & $(\mathbf{6},\, \mathbf{15}) \oplus (\mathbf{10},\, \mathbf{10})$\\
         $\overline{\mathbf{190}}$ & $(\mathbf{6},\, \overline{\mathbf{15}}) \oplus (\overline{\mathbf{10}},\, \overline{\mathbf{10}})$\\
         $\mathbf{210}$ & $(\mathbf{10},\, \mathbf{15}) \oplus (\mathbf{6},\, \mathbf{10})$\\
         $\overline{\mathbf{210}}$ & $(\overline{\mathbf{10}},\, \overline{\mathbf{15}}) \oplus (\mathbf{6},\, \overline{\mathbf{10}})$\\
         $\mathbf{399}$ & $(\mathbf{15},\, \mathbf{24})\oplus (\mathbf{15},\, \mathbf{1})\oplus (\mathbf{1},\, \mathbf{24})$\\
       \hline
    \end{tabular}
    \caption{Branching rules of special embedding $su(20)\supset su(4)\oplus su(5)$~\cite{Yamatsu:2015npn}.}
    \label{tab:su20}
\end{table}

\section{Alternative realizations of special embedding}\label{sec:alternative}

In the main text of the paper, we have specially embedded the original composite axion model into a $SU(4N)\times SU(3)_2\times SU(N)_2 \times U(1)_X$ gauge theory to solve the domain wall problem. Besides from the model shown in section~\ref{sec:model}, we here present some alternative methods to realize the special embedding.

\subsection{Tri-fundamental representation}

\begin{table}[t!]
    \centering
    \renewcommand{\arraystretch}{1.5}
    \setlength{\tabcolsep}{1.5pt}
    \begin{tabular}{|c|c|c|c|c|c|c|c|c|c|}
        \hline
        Fields &  $[SU(4N)]$  & $[SU(3)_2]$ & $[SU(N)_2]$  & $[U(1)_{X}]$ & $U(1)_{A}$  & $U(1)_{B}$ &  $SU(4N)_{L}$ & $SU(4N)_{R}$  \\ \hline
       $\Psi$&    $\mathbf{4N}$ & $\mathbf{1}$ & $\mathbf{1}$ & $x$ & 1 & 1 &  $\mathbf{4N}$ & $\mathbf{1}$ \\ \hline
       $\bar\Psi$ & $\overline{\mathbf{4N}}$ & $\mathbf{1}$ & $\mathbf{1}$ & $-x$ & 0  & $-1$ &  $\mathbf{1}$ & $\overline{\mathbf{4N}}$\\ \hline
       $\Phi$ & $\mathbf{4N}$ & $\overline{\mathbf{3}}$ & $\overline{\mathbf{N}}$ & $-z$ & 0 & 0 &  $\mathbf{1}$ & $\mathbf{1}$\\ \hline
    \end{tabular}
    \vspace{3mm}
    \caption{Matter content for the model with the tri-fundamental representation.}
    \label{tab:special1_Y}
\end{table}

Let us consider a 
$SU(4N)\times SU(3)_2\times SU(N)_2$ gauge theory.
The matter content of this model is summarized in Table~\ref{tab:special1_Y}.
The model contains Weyl fermions $\Psi,\bar\Psi$ and a complex scalar field $\Phi$.
The SM quarks are charged under $[SU(3)_2]$, so that after the gauge symmetry breaking, they are charged under $[SU(3)_c]$.
The diagonal subgroup of the $SU(4N)_L\times SU(4N)_R$ flavor symmetry is gauged, i.e. $[SU(4N)]$.
The SM $U(1)_Y$ symmetry is the linear combination of $U(1)_X$ and the $U(1)$ coming from the symmetry breaking of $SU(4) (\subset SU(4N)) \to SU(3)\times U(1)$. The SM fermions are charged under $U(1)_X$ to have their hypercharges.

We can obtain the gauge symmetry breaking,
\begin{align}
    SU(4N)\times SU(3)_2\times SU(N)_2 \times U(1)_X \to SU(3)_c \times U(1)_Y \times SU(N)_V \ ,
\end{align}
by a VEV of $\Phi$, which is expressed as a $4N\times 3N$ matrix,
\begin{align}
    \langle\Phi\rangle= {\rm diag}_N\left( \begin{pmatrix}
    v & 0 & 0 \\
    0 & v & 0 \\
    0 & 0 & v \\
    0 & 0 & 0
\end{pmatrix}, ... ,\begin{pmatrix}
    v & 0 & 0 \\
    0 & v & 0 \\
    0 & 0 & v \\
    0 & 0 & 0
\end{pmatrix} \right)  \ ,
\end{align}
where $v$ is a parameter with mass dimension one.
Under $SU(N)_V\times SU(3)_c \times U(1)_Y$, the fermions $\Psi$ and $\bar\Psi$ transform as
\begin{align}
    \Psi:(\mathbf{N},\mathbf{3},z+x)\oplus (\mathbf{N},\mathbf{1},-3z+x) \ , \quad \bar \Psi:(\bar{\mathbf{N}},\bar{\mathbf{3}},-z-x)\oplus (\bar{\mathbf{N}},\mathbf{1},3z-x)\ ,
\end{align}
which is the same as the original composite axion model after identifying $a=z+x$ and $b=-3z+x$.

\subsection{Large representation}

\begin{table}[t!]
    \centering
    \renewcommand{\arraystretch}{1.5}
    \setlength{\tabcolsep}{1.5pt}
    \begin{tabular}{|c|c|c|c|c|c|c|c|c|}
        \hline
        Fields &  $[SU(4N)]$  &  $[SU(3)]$ & $[U(1)_{X}]$  & $U(1)_{A}$  & $U(1)_{B}$ &  $SU(4N)_{L}$ & $SU(4N)_{R}$  \\ \hline
       $\Psi$&    $\mathbf{4N}$ & $\mathbf{1}$ & $x$ & 1  & 1 &  $\mathbf{4N}$ & $\mathbf{1}$ \\ \hline
       $\bar\Psi$ & $\overline{\mathbf{4N}}$ & $\mathbf{1}$ & $-x$ & 0 & $-1$ &  $\mathbf{1}$ & $\overline{\mathbf{4N}}$ \\ \hline
       $\Phi$ & $\mathbf{r}$ & $\mathbf{1}$ & 0 & 0 & 0 &  $\mathbf{1}$ & $\mathbf{1}$ \\ \hline
       $\Phi'$ & $\mathbf{r'}$ & $\mathbf{r_3}$ & $z$ & 0 & 0 &  $\mathbf{1}$ & $\mathbf{1}$ \\ \hline
    \end{tabular}
    \vspace{3mm}
    \caption{Matter content for the model with a large representation.
    }
    \label{tab:special3}
\end{table}

Another method is to spontaneously break $SU(4N)$ directly into $SU(N)\times SU(4)$. The gauge symmetry of the model can be then reduced to $SU(4N)\times SU(3)$. However, this requires a very large representation $\mathbf{r}$ that contains a singlet at the direction of the focused symmetry breaking. The matter content is summarized in Table~\ref{tab:special3}. 
The model contains Weyl fermions $\Psi,\bar\Psi$ and complex scalar fields $\Phi,\Phi'$.
The SM quarks are charged under the $[SU(3)]$, so that after the gauge symmetry breaking, they are charged under the ordinary color $[SU(3)_c]$.

For each $N$, we can find the representation $\mathbf{r}$ responsible for the symmetry breaking as follows~\cite{Yamatsu:2015npn},
\begin{itemize}
    \item $\mathbf{r} = \mathbf{35700}$ for $SU(4N=20)$ ($N=5$)\\[1ex]
    Under the symmetry breaking of $SU(20)\to SU(5)\times SU(4)$, the large representation is decomposed as \\[0.5ex]
    $\mathbf{35700} = (\mathbf{200},\mathbf{20}') \oplus (\mathbf{200},\mathbf{15}) \oplus (\mathbf{200},\mathbf{1}) \oplus (\mathbf{126},\overline{\mathbf{45}}) \oplus (\mathbf{126},\mathbf{15}) \oplus (\overline{\mathbf{126}},\mathbf{45}) \oplus (\overline{\mathbf{126}},\mathbf{15}) \oplus (\mathbf{75},\mathbf{84}) \oplus (\mathbf{75},\mathbf{15}) \oplus (\mathbf{75},\mathbf{1}) \oplus (\mathbf{24},\mathbf{84}) \oplus (\mathbf{24},\mathbf{45}) \oplus (\mathbf{24},\overline{\mathbf{45}}) \oplus (\mathbf{24},\mathbf{20}') \oplus 3(\mathbf{24},\mathbf{15}) \oplus (\mathbf{24},\mathbf{1}) \oplus (\mathbf{1},\mathbf{84}) \oplus (\mathbf{1},\mathbf{20}') \oplus (\mathbf{1},\mathbf{15}) \oplus (\mathbf{1},\mathbf{1})$~,
    \item $\mathbf{r} = \mathbf{3876}$ or $\overline{\mathbf{3876}}$ for $SU(4N=16)$ ($N=4$)\\[1ex]
    Under the symmetry breaking of $SU(16)\to SU(4)\times SU(4)$, the large representation is decomposed as \\[0.5ex]
    $\mathbf{3876}=(\mathbf{35},\mathbf{35}) \oplus (\mathbf{45},\mathbf{45}) \oplus (\mathbf{20}',\mathbf{20}') \oplus (\mathbf{15},\mathbf{15}) \oplus (\mathbf{1},\mathbf{1})$~,
    \item $\mathbf{r} = \mathbf{4212}$ for $SU(4N=12)$ ($N=3$)\\[1ex]
    Under the symmetry breaking of $SU(12)\to SU(3)\times SU(4)$, the large representation is decomposed as \\[0.5ex]
    $\mathbf{4212} = (\mathbf{8},\mathbf{84}) \oplus (\mathbf{1},\mathbf{84}) \oplus (\overline{\mathbf{10}},\mathbf{45}) \oplus (\mathbf{8},\mathbf{45}) \oplus (\mathbf{10},\overline{\mathbf{45}}) \oplus (\mathbf{8},\overline{\mathbf{45}}) \oplus (\mathbf{27},\mathbf{20}') \oplus (\mathbf{8},\mathbf{20}') \oplus (\mathbf{1},\mathbf{20}') \oplus (\mathbf{27},\mathbf{15}) \oplus (\mathbf{10},\mathbf{15}) \oplus (\overline{\mathbf{10}},\mathbf{15}) \oplus 3(\mathbf{8},\mathbf{15}) \oplus (\mathbf{1},\mathbf{15}) \oplus (\mathbf{27},\mathbf{1}) \oplus (\mathbf{8},\mathbf{1}) \oplus (\mathbf{1},\mathbf{1})$~.
\end{itemize}
We can achieve $SU(4N) \xrightarrow{\langle\Phi\rangle} SU(N)\times SU(4)$ without an additional $SU(N)$ gauge group.


\bibliography{reference}
\bibliographystyle{JHEP}

\end{document}